\documentclass[a4paper,12pt]{article}

\usepackage{amsmath}\usepackage{amssymb}
\usepackage{latexsym}\usepackage{cite}

\usepackage{amsmath}
\usepackage[dvips]{graphicx}
\usepackage{pstricks}
\usepackage{color}
\usepackage{bm}

\usepackage{graphicx}

\definecolor{Red}{rgb}{1.00, 0.00, 0.00}
\definecolor{Green}{rgb}{0.00, 1.00, 0.00}
\definecolor{Blue}{rgb}{0.00, 0.00, 1.00}
\definecolor{Cyan}{rgb}{0.00, 1.00, 1.00}
\definecolor{Mymagenta}{rgb}{0.3, 0.00, 1.00}%
\definecolor{Magenta}{rgb}{1.00, 0.00, 1.00}
\definecolor{DeepSkyBlue}{rgb}{0.00, 0.75, 1.00}
\definecolor{DarkGreen}{rgb}{0.00, 0.39, 0.00}
\definecolor{SpringGreen}{rgb}{0.00, 1.00, 0.50}
\definecolor{Mygreen}{rgb}{0.00, 0.72, 0.00}
\definecolor{DarkOrange}{rgb}{1.00, 0.55, 0.00}
\definecolor{OrangeRed}{rgb}{1.00, 0.27, 0.00}
\definecolor{DeepPink}{rgb}{1.00, 0.08, 0.57}
\definecolor{DarkViolet}{rgb}{0.58, 0.00, 0.82}
\definecolor{SaddleBrown}{rgb}{0.57, 0.27, 0.07}
\definecolor{Black}{rgb}{1.00, 1.00, 1.00}
\definecolor{Ablue}{rgb}{0.10, 0.1, 1.00}

\newcommand{\be}{\begin{equation}}
\newcommand{\ee}{\end{equation}}
\newcommand{\tr}{\,\textup{tr}}

\def\F{{\bf F}}

\def\beq{\begin{equation}}
\def\eeq{\end{equation}}
\def\beqr{\begin{eqnarray}}
\def\eeqr{\end{eqnarray}}
\def\pl{\partial}

\def\al{\alpha}
\def\bt{\beta}
\def\Ga{\Gamma}
\def\ga{\gamma}
\def\de{\delta}
\def\De{\Delta}

\def\si{\sigma}
\def\Si{\Sigma}
\def\te{\theta}

\def\La{\Lambda}
\def\lam{\lambda}
\def\Om{\Omega}
\def\om{\omega}
\def\ep{\epsilon}

\def\sq{\sqrt}

\def\l{\left (}
\def\r{\right )}

\def\fr{\frac}
\def\la{\label}
\def\hs{\hspace}
\def\vs{\vspace}

\def\ran{\rangle}
\def\lan{\langle}
\def\ov{\overline}
\def\tl{\tilde}
\def\tm{\times}
\def\larr{\leftarrow}
\def\rarr{\rightarrow}

\def\lrarr{\longrightarrow}

\begin{document}

\begin{flushright}
CETUP2013-013\\
March 1, 2014 \\
\end{flushright}

\vs{1.5cm}

\begin{center}
{\Large\bf

$SU(5)\tm SU(5)'$ unification and $D_2$ parity:\\
 Model for composite leptons}

\end{center}

\vspace{0.5cm}
\begin{center}
{\large
{}~Zurab Tavartkiladze\footnote{E-mail: zurab.tavartkiladze@gmail.com}
}
\vspace{0.5cm}

{\em Center for Elementary Particle Physics, ITP, Ilia State University, 0162 Tbilisi, Georgia}

\end{center}
\vspace{0.6cm}

\begin{abstract}


We study a grand unified $SU(5)\tm SU(5)'$ model supplemented by $D_2$ parity. The $D_2$ greatly reduces the number of parameters
and is important for phenomenology. The model, we present, has various novel and interesting properties.
 Because of the specific pattern of grand unification symmetry breaking and emerged strong dynamics at low energies, the Standard Model leptons, along
with right-handed/sterile neutrinos, come out as composite states.
The generation of the charged fermion and neutrino masses are studied within the considered
scenario. Moreover, the issues of gauge coupling unification and nucleon stability are investigated in details.
Various phenomenological implications  are also discussed.

\end{abstract}

\section{Introduction}

The Standard Model (SM) of electroweak interactions has been a very successful theory for decades.
The triumph of this celebrated model occurred thanks to the Higgs boson discovery \cite{Aad:2012tfa} at CERN's Large Hadron Collider.
In spite of this success, several phenomenological and theoretical issues motivate one to think of some physics beyond the SM.
 Because of renormalization running, the self-coupling of the SM Higgs boson becomes negative at scale near $\sim 10^{10}$~GeV
\cite{Buttazzo:2013uya}, \cite{Sher:1988mj}
(with the  Higgs mass$\simeq 126$~GeV), causing vacuum instability (becoming more severe within the inflationary setup; see the discussion in Sec. \ref{impl}).
Moreover, the SM fails to accommodate atmospheric and solar neutrino data \cite{nu-data}. The renormalizable part of the SM interactions
render neutrinos to be massless. Also, Planck scale suppressed $d=5$ lepton number violating operators do not
generate neutrino mass with desirable magnitude.
These are already strong motivations to think about the existence of some new physics between electroweak (EW) and Planck scales.

Among various extensions of the SM, the grand unification (GUT) \cite{Pati:1974yy},  \cite{GG-gut} is a leading candidate.
Unifying all gauge interactions in a single group, at high
energies one can deal with a single unified gauge coupling. At the same time, quantization of quark and lepton charges occurs by embedding
all fermionic states in unified GUT multiplets. The striking prediction of the grand unified theory is the baryon number violating nucleon decay.
This opens the prospect for probing the nature at very short distances.
GUTs based on $SO(10)$ symmetry \cite{Fritzsch:1974nn} [which includes $SU(2)_L\tm SU(2)_R\tm SU(4)_c$ symmetry \cite{Pati:1974yy} as a maximal subgroup]
involve right-handed neutrinos (RHNs), which provide a simple and elegant way for neutrino mass generation via the seesaw mechanism \cite{seesaw}.
In spite of these salient futures,  GUT model building  encounters numerous problems and phenomenological
difficulties. With single scale breaking, i.e., with no new interactions and/or intermediate states between EW and GUT scales, grand unified theories
[such as minimal $SU(5)$ and $SO(10)$] do not lead to successful gauge coupling unification.
Besides this,  building GUT with the realistic fermion sector, understanding the GUT symmetry breaking pattern, and avoiding too rapid nucleon decay remain a
great challenge.

Motivated by these issues, we consider $SU(5)\tm SU(5)'$ GUT augmented with $D_2$ parity (exchange symmetry).
The latter, relating two $SU(5)$ gauge groups, reduces the number of parameters, and at and above the GUT scale, one deals with single
gauge coupling. The grand unified theories with $SU(5)\tm SU(5)'$ symmetry, considered in earlier works \cite{1product-group-GUT}, in which
at least one gauge factor of the SM symmetry emerges as a diagonal subgroup,
have been proven to be very successful for building models with realistic phenomenology. However, to our knowledge,  in such constructions the  $D_2$ parity
has not been applied before.\footnote{In the second citation of Ref. \cite{1product-group-GUT}, the exchange symmetry
 was considered; however, some terms violating this symmetry have been included.} The reason could be the prejudice of remaining with extra unwanted chiral matter states in the spectrum.
However, within our model due to specific construction, this does not happen, and  below the few-TeV scale, surviving states are just of
the Standard Model.
The  $D_2$ parity also plays a crucial role for phenomenology and has interesting implications.
By the specific pattern of the $SU(5)\tm SU(5)'$ symmetry breaking and spectroscopy, the successful gauge coupling unification is obtained. Interestingly, within the considered framework,
the SM leptons emerge as a composite states,
 while the quarks are fundamental objects.
Lepton mass generation occurs by a new mechanism, finding natural realization within a presented
model. Since leptons and quarks have different footing, there is no problem of their mass degeneracy
(unlike the minimal SO(10) and $SU(5)$ grand unified theories, which require some extensions \cite{Ellis:1979fg}).
 Moreover, along with composite SM leptons, the model involves three families of composite SM singlet fermionic states, which may be identified with RHNs or sterile neutrinos.  Thus, the neutrino masses can be generated. In addition, we show that, due to the specific fermion pattern,  $d=6$ nucleon decay
 can be adequately suppressed within the considered model.
 The model also has various interesting properties and implications, which we also discuss.
 Since two $SU(5)$ groups will be related by $D_2$ parity, initial states will be doubled, i.e., will be introduced in twins. Because of
 this, we refer to the proposed $SU(5)\tm SU(5)'\tm D_2$ model as twinification.

The paper is organized as follows.
In the next section, first we introduce the $SU(5)\tm SU(5)'\tm D_2$ GUT and discuss the symmetry breaking pattern. Then, we present the
spectrum of bosonic states.   In Sec. \ref{ferm}, considering the fermion sector, we give transformation properties of the GUT matter multiplets under
$D_2$ parity and build the Yukawa interaction Lagrangian. The latter is responsible for the generation of quark masses and CKM matrix elements. Because of the specific
pattern of the symmetry breaking and strong $SU(3)'$ [originating from $SU(5)'$ gauge symmetry] dynamics, the SM leptons emerge as composite
objects. We present a novel mechanism for composite lepton mass generation. Together with the SM leptons, three families of right-handed/sterile
neutrinos are composite. We also discuss the neutrino mass generation within our scenario. In Sec. \ref{unif} we give details of gauge coupling unification.
The issue of nucleon stability is addressed in Sec. \ref{Pdecay}.  Although the GUT scale, within our model, comes out to be relatively low
($\simeq \!5\cdot 10^{11}$~GeV), we show that the $d=6$ baryon number violating operators can be adequately suppressed.
This happens to be possible due to the specific pattern of the fermion sector we are suggesting. In Sec. \ref{impl} we summarize and discuss various phenomenological constraints and possible
implications of the considered scenario. We also emphasize the model's peculiarities and novelties, which open broad prospects for further investigations.
Appendix \ref{comp-anom} discusses details related to the compositeness and anomaly matching conditions.
In Appendix \ref{rg} we give details of the gauge coupling unification. In particular, the renormalization group (RG) equations and $b$ factors at various
energy intervals are presented. The short-range renormalization of baryon number violating $d=6$ operators is also performed.

\section{$SU(5)\tm SU(5)'\tm D_2$ Twinification}
\la{55model}

Let us consider the theory based on $SU(5)\tm SU(5)'$ gauge symmetry. Besides this symmetry, we postulate discrete parity $D_2$, which exchanges
two $SU(5)$'s. Therefore, the symmetry of the model is
\beq
G_{GUT}=SU(5)\tm SU(5)'\tm D_2~.
\la{model-sym}
\eeq
As noted, the action of $D_2$ interchanges the gauge fields (in adjoint representations) of $SU(5)$ and $SU(5)'$,
\beq
D_2~:~~~~(A_{\mu})^a_b\to (A_{\mu}')^{a'}_{b'}~,~~~(A_{\mu}')^{a'}_{b'}\to (A_{\mu})^a_b~,
\la{D5gauge}
\eeq
with $(A_{\mu})^{a}_{b}=\fr{1}{2}\sum_{i=1}^{24}A_{\mu}^{i}(\lam^i)^{a}_{b}$ and
$(A_{\mu}')^{a'}_{b'}=\fr{1}{2}\sum_{i'=1}^{24}A_{\mu}^{'i'}(\lam^{i'})^{a'}_{b'}$,
where $a,b$ and $a',b'$ denote indices of $SU(5)$ and $SU(5)'$ respectively. The $\lam^i, \lam^{i'}$ are corresponding Gell-Mann matrices.
Thanks to the $D_2$, at and above the GUT scale $M_G$, we have single gauge coupling
\beq
{\al_5}=\al_{5'}~.
\la{g-unif}
\eeq
 Grand unified theories based on product groups allow us to build simple models with realistic phenomenology
\cite{1product-group-GUT}, \cite{2product-group-GUT}. In our case, as we show below, the EW part [i.e., $SU(2)_w\tm U(1)_Y$] of the SM gauge symmetry
will belong to the diagonal subgroup of $SU(5)\tm SU(5)'$.

\vspace{0.4cm}
{\bf Potential and symmetry breaking}
\vspace{0.15cm}

For $G_{GUT}$ symmetry  breaking and building realistic phenomenology,
we introduce the states
\beq
H\sim (5,1)~,~~~\Si \sim (24,1)~,~~~~~H'\sim (1,5)~,~~~\Si' \sim (1,24)~,~~~
\Phi \sim (5, \bar 5)~,
\la{sc-gauge-tr}
\eeq
where in brackets transformation properties under $SU(5)\tm SU(5)'$ symmetry are indicated. $H$ includes SM Higgs doublet $h$. The introduction of $H'$
is required by $D_2$ symmetry. By the same reason, two adjoints $\Si$ and $\Si'$  (needed for GUT symmetry breaking) are introduced.
The bifundamental state  $\Phi$ will also  serve for desirable symmetry breaking.

The action of $D_2$ parity on these fields is
\beq
D_2~:~~~~~~H_a \stackrel{\rarr}{_\larr} H_{a'}'~,~~~~\Si^a_b \stackrel{\rarr}{_\larr} {\Si'}^{a'}_{b'}~,~~~~
\Phi_a^{b'}\stackrel{\rarr}{_\larr} (\Phi^\dag)_{a'}^b ~,
\la{sc-D2-tr}
\eeq
where we have made explicit the indices of $SU(5)$ and $SU(5)'$ . With  Eqs. (\ref{sc-D2-tr}), (\ref{D5gauge}) and  (\ref{g-unif}) one can easily make sure that  the kinetic part
$|D_{\mu}H|^2+|D_{\mu}H'|^2+\fr{1}{2}\tr (D_{\mu}\Si )^2+\fr{1}{2}\tr (D_{\mu}\Si' )^2+|D_{\mu}\Phi|^2$
 of
the scalar field Lagrangian is invariant.

The scalar potential, invariant under $G_{GUT}$ symmetry [of Eq. (\ref{model-sym})] is
\beq
V=V_{H\Si}+V_{H'\Si'}+V_{mix}^{(1)}+V_{\Phi}+V_{mix}^{(2)}~,
\la{V-tot}
\eeq
with
$$
V_{H\Si}=-M_{\Si}^2{\tr}\Si^2+\lam_1({\tr}\Si^2)^2+\lam_2{\tr}\Si^4+H^\dag \l M_H^2-h_1\Si^2+h_2{\tr}\Si^2\r H
+\lam_H(H^\dag H)^2~,
$$
$$
V_{H'\Si'}=-M_{\Si}^2{\tr}{\Si'}^2\!+\!\lam_1({\tr}{\Si'}^2)^2\!\!+\!\lam_2{\tr}{\Si'}^4\!+\!{H'}^\dag \!\l M_H^2\!-\!h_1{\Si'}^2
\!+\!h_2{\tr}{\Si'}^2\r \!\!H'
\!+\!\lam_H({H'}^\dag H')^2~,
$$
$$
V_{mix}^{(1)}=\lam ({\tr}\Si^2)({\tr}{\Si'}^2)+\tl{h} \l  H^\dag H{\tr}{\Si'}^2+{H'}^\dag H'{\tr}\Si^2 \r
+\hat h(H^\dag H)({H'}^\dag H')~,
$$
$$
V_{\Phi}=-M_{\Phi}^2\Phi^\dag \Phi+\lam_{1\Phi}\l \Phi^\dag \Phi\r^2+\lam_{2\Phi}\Phi^\dag \Phi \Phi^\dag \Phi~,
$$
$$
V_{mix}^{(2)}=\mu (H^\dag \Phi H'\!+\!H\Phi^\dag {H'}^\dag )
\!+\!\fr{\lam_{1H\Phi }}{\sqrt{25}}(\Phi^\dag \Phi)\left [(H^\dag H)\!+\!({H'}^\dag H')\right ]\!+\!
\fr{\lam_{2H\Phi }}{\sqrt{10}}\l  \!H^\dag \Phi  \Phi^\dag \!H \!+\! {H'}^\dag \Phi^\dag \Phi \!H' \!\r +
$$
\beq
\lam_{1\Si \Phi}(\Phi^\dag \Phi)({\tr}\Si^2+{\tr}{\Si'}^2)-\lam_{2\Si \Phi}(\Phi^\dag \Si^2\Phi +\Phi {\Si'}^2\Phi^\dag )~.
\la{V-terms}
\eeq
To make analysis simpler, we have omitted terms with first powers of  $\Si $ and $\Si'$ (such as $H^\dag\Si H$, ${H'}^\dag\Si' H'$, etc.) and
also cubic terms of  $\Si $ and $\Si'$. This simplification can be achieved by  $Z_2$ discrete symmetry and will not harm anything.

The potential terms and couplings in Eqs. (\ref{V-tot}) and (\ref{V-terms}) allow us to have a desirable
and self-consistent pattern of symmetry breaking.
First, we will sketch the symmetry breaking pattern.  Then, we will analyze  the potential and discuss the spectrum of
bosonic states.
 We will stick to several stages of the GUT symmetry breaking. At the first step, the $\Si $ develops
the vacuum expectation value (VEV)$\sim M_G$ with
\beq
\lan \Si \ran =v_{\Si}{\rm Diag}\l 2, 2, 2, -3, -3\r ~,~~~~v_{\Si}\sim M_G~.
\la{Si-VEV}
\eeq
This causes the symmetry breaking:
\beq
SU(5)\stackrel{\lan \Si \ran }{_{\lrarr}} SU(3)\tm SU(2)\tm U(1)\equiv G_{321} .
\la{stage1}
\eeq
We select VEVs of $\Si' $ and $\Phi$ much smaller than $M_G$. As it will turn out, the phenomenologically preferred scenario is
 $\lan \Si' \ran \sim 4\cdot 10^{6}$~GeV and $\lan \Phi\ran \sim 8\cdot 10^{4}$~GeV.
 With
 \beq
\lan \Si' \ran =v_{\Si'}{\rm Diag}\l 2, 2, 2, -3, -3\r ~,
\la{Si1-VEV}
\eeq
the breaking
\beq
SU(5)'\stackrel{\lan \Si' \ran }{_{\lrarr}} SU(3)'\tm SU(2)'\tm U(1)'\equiv {G_{321}}'
\la{stage2}
\eeq
is achieved.
The last stage of the GUT breaking is done by $\lan \Phi\ran $ with a direction
\beq
\lan \Phi \ran =v_{\Phi }\cdot {\rm Diag }\l 0, ~0,~ 0,~ 1,~ 1\r .
\la{vev-conf}
\eeq
This configuration of $\lan \Phi \ran$ breaks symmetries  $SU(2)\tm U(1)$ [subgroup of $SU(5)$] and $SU(2)'\tm U(1)'$ [subgroup of $SU(5)'$] to the diagonal symmetry group:
\beq
SU(2)\tm U(1)\tm SU(2)'\tm U(1)'\stackrel{\lan \Phi \ran }{_{\lrarr}} \left [ SU(2)\tm U(1)\right ]_{\rm diag}~.
\la{Phi-br}
\eeq
As we see, all VEVs preserve $SU(3)$ and $SU(3)'$ groups arising from $SU(5)$ and
$SU(5)'$ respectively. However, unbroken $SU(2)_{\rm diag}$ is coming (as superposition) partly from $SU(2)\subset SU(5)$
and partly from $SU(2)'\subset SU(5)'$. Similar applies to $U(1)_{\rm diag}$; i.e., it is superposition of two Abelian factors:
$U(1)\subset SU(5)$ and $U(1)'\subset SU(5)'$.

Now,  making the identifications
\beq
SU(3)\equiv SU(3)_c~,~~~~~SU(2)_{\rm diag}\equiv SU(2)_w~,~~~~~U(1)_{\rm diag}\equiv U(1)_Y
\la{identif}
\eeq
and taking into account Eqs. (\ref{stage1}), (\ref{stage2}), and  (\ref{Phi-br}), we can see that GUT symmetry is broken as:
\beq
 G_{GUT}\to SU(3)_c\tm SU(2)_w\tm U(1)_Y\tm SU(3)'=G_{SM}\tm SU(3)'~,
\la{final-br}
\eeq
where $G_{SM}=SU(3)_c\tm SU(2)_w\tm U(1)_Y$ denotes the SM gauge symmetry. Because of these, at the intermediate scale $\mu =M_I (\sim \lan \Phi\ran )$, we will have
the matching conditions for the gauge couplings,
\beq
{\rm at}~\mu=M_I:~~~~~~\fr{1}{g_w^2}=\fr{1}{g_2^2}+\fr{1}{{g}_{2'}^2} ~,~~~~
~~~\fr{1}{g_Y^2}=\fr{1}{g_1^2}+\fr{1}{{g}_{1'}^2}~,
\la{g-match-MI}
\eeq
where subscripts indicate  to which gauge interaction the appropriate coupling corresponds [e.g., $g_{1'}$ is the coupling of $U(1)'$ symmetry, etc.].

The extra $SU(3)'$ factor has important and interesting implications, which we discuss below.

As was mentioned, while $\lan \Si \ran \sim M_G$, the VEVs $\lan \Phi \ran $ and $\Si'$ are at intermediate scales $M_I$ and ${M_I}'$, respectively,
\beq
v_{\Phi}\sim M_I,~~~~v_{\Si'}\sim {M_I}' ~,
\la{MI-MI1}
\eeq
with the hierarchical pattern
\beq
M_I\ll {M_I}'\ll M_G~.
\la{VEV-hier}
\eeq

Detailed analysis of the whole potential shows that there is true minimum along directions (\ref{Si-VEV}), (\ref{Si1-VEV}), and
(\ref{vev-conf}) with  $\lan H\ran =\lan H'\ran =0$.
With $\lan \Si \ran \neq \lan \Si' \ran$, the $D_2$ is broken spontaneously. The residual $SU(3)'$ symmetry  will play an important role,
and the hierarchical pattern of Eq. (\ref{VEV-hier}) will turn out to be crucial for successful gauge coupling unification (discussed below).

The hierarchical pattern   (\ref{VEV-hier}), of the GUT symmetry breaking, makes it simple to minimize the potential
and analyze the spectrum.

Three extremum conditions, determining $v_{\Si}, v_{\Si'}$ and $v_{\Phi}$ along the directions (\ref{Si-VEV}), (\ref{Si1-VEV}) and (\ref{vev-conf})
and obtained from whole potential, are
$$
10(30\lam_1+7\lam_2)v_{\Si}^2+150\lam v_{\Si'}^2+(10\lam_{1\Si \Phi}-3\lam_{2\Si \Phi})v_{\Phi }^2=5M_{\Si}^2 ~,
$$
$$
150\lam v_{\Si }^2+10(30\lam_1+7\lam_2)v_{\Si'}^2+(10\lam_{1\Si \Phi}-3\lam_{2\Si \Phi})v_{\Phi }^2=5M_{\Si}^2~,
$$
\beq
3(10\lam_{1\Si \Phi}-3\lam_{2\Si \Phi})(v_{\Si }^2+v_{\Si' }^2)+(4\lam_{1\Phi}+2\lam_{2\Phi})v_{\Phi }^2=M_{\Phi}^2~.
\la{min-conditions}
\eeq
Because of hierarchies (\ref{MI-MI1}) and (\ref{VEV-hier}), from the first equation of Eq. (\ref{min-conditions}), with a good approximation we obtain
\beq
v_{\Si}\simeq \fr{M_{\Si}}{\sqrt{2(30\lam_1+7\lam_2)}}~.
\la{Si-vev-sol}
\eeq
Thus, with $2(30\lam_1+7\lam_2)\sim 1$, we should have $M_{\Si}\approx M_G$.
On the other hand, from the last two equations of Eq. (\ref{min-conditions}), we derive
\beq
v_{\Si'}^2\simeq \fr{M_{\Si}^2-30\lam v_{\Si}^2}{2(30\lam_1+7\lam_2)}~,~~~
v_{\Phi}^2=\fr{M_{\Phi}^2-3(10\lam_{1\Si \Phi}-3\lam_{2\Si \Phi})(v_{\Si}^2+v_{\Si'}^2)}{4\lam_{1\Phi}+2\lam_{2\Phi}}~.
\la{Si1-Phi-vev-sol}
\eeq
To obtain the scales $M_I$ and ${M_I}'$, according to Eqs. (\ref{MI-MI1}) and (\ref{VEV-hier}), we have to arrange (by price of tunings)
$M_{\Si}^2-30\lam v_{\Si}^2\approx ({M_I}')^2$ and
$M_{\Phi}^2-3(10\lam_{1\Si \Phi}-3\lam_{2\Si \Phi})(v_{\Si}^2+v_{\Si'}^2)\approx M_I^2$ [with $(4\lam_{1\Phi}+2\lam_{2\Phi})\sim 1$].

\vspace{0.4cm}
{\bf The Spectrum}
\vspace{0.15cm}

At the first stage of symmetry breaking, the $(X,Y)$ gauge bosons [of $SU(5)$] obtain GUT scale masses. They absorb appropriate states (with quantum numbers of leptoquarks)
from the adjoint scalar $\Si$.
The remaining physical fragments $(\Si_8, \Si_3, \Si_1)$ [the $SU(3)$ octet, $SU(2)$ triplet, and a singlet, respectively]
receive GUT scale masses.  These states are heaviest and their mixings with other ones can be neglected. From Eq. (\ref{V-terms}), with Eq. (\ref{min-conditions})
we get
\beq
M_{\Si_8}^2\simeq 20\lam_2v_{\Si}^2~,~~~M_{\Si_3}^2\simeq 80\lam_2v_{\Si}^2~,~~~M_{\Si_1}^2\simeq 4M_{\Si}^2~.
\la{Si-spectrum}
\eeq
Further, we will not give masses of states that are singlets under all symmetry groups.
The mass square of the $SU(3)'$ octet (from $\Si'$) is
\beq
M_{{\Si'}_{8'}}^2=20\lam_2v_{\Si'}^2+\fr{6}{5}\lam_{2\Si \Phi }v_{\Phi }^2~.
\la{Si1-81-mass}
\eeq
The triplet ${\Si'}_{3'}$ mixes with a real (CP even) $SU(2)_w$ triplet $\Phi_3$ (from $\Phi $). [Both these states are real adjoints of $SU(2)_w$.] The
appropriate mass squared couplings are
\beq
\fr{1}{2} \l {\Si^{'i}}_{3'}, ~ \Phi_3^i\r \!
\left( \!
  \begin{array}{cc}
    4M_{\Si'_{8'}}^2\!-\!\fr{28}{5}\lam_{2\Si \Phi }v_{\Phi }^2 & 6\sqrt{2}\lam_{2\Si \Phi }v_{\Phi }v_{\Si'} \\
   6\sqrt{2}\lam_{2\Si \Phi }v_{\Phi }v_{\Si'} & 4\lam_{2\Phi}v_{\Phi}^2 \\
  \end{array}
\! \right)\!\!
\l \!\!\!
  \begin{array}{c}
    {\Si^{'i}}_{3'} \\
    \Phi_3^i \\
  \end{array}\!\!\!\r ,
\la{triplet-mixing}
\eeq
where $i=1,2,3$ labels the components of the $SU(2)_w$ adjoint.
The CP-odd real $SU(2)_w$ triplet from   $\Phi $ is absorbed by appropriate gauge fields after $SU(2)\tm SU(2)'\to SU(2)_w$ breaking and becames genuine
Goldstone modes.

By the VEVs $v_{\Si}$ and $v_{\Si'}$, the symmetry $SU(5)\tm SU(5)'\tm D_2$ is broken down to
 $G_{321}\tm {G_{321}}'$ [see Eqs. (\ref{stage1}) and (\ref{stage2})]. Thus, between the scales $M_I$ and ${M_I}'$,
 we have this symmetry, and the $\Phi(5, \bar 5)$ splits into fragments
 \beq
\Phi(5, \bar 5)=\Phi_{DD'}\oplus \Phi_{DT'} \oplus \Phi_{TT'} \oplus  \Phi_{TD'}
\la{Phi55-dec}
\eeq
with transformation properties under $G_{321}\tm {G_{321}}'$ given by
\begin{eqnarray}
G_{321}\tm {G_{321}}'~:&~~~\Phi_{DD'} \sim \l 1, 2, -\fr{3}{\sq{60}}, 1,2', \fr{3}{\sq{60}}\r ,&
\Phi_{DT'}\sim \l 1, 2, -\fr{3}{\sq{60}}, \bar 3',1, -\fr{2}{\sq{60}}\r ,\nonumber \\
&~~ \Phi_{TT'} \sim \l 3, 1, \fr{2}{\sq{60}}, \bar 3',1, -\fr{2}{\sq{60}}\r ,&
\Phi_{TD'}\sim \l 3, 1, \fr{2}{\sq{60}}, 1,2', \fr{3}{\sq{60}}\r .
\la{Phi-frag-trans}
\end{eqnarray}
The masses of these fragments will be denoted by $M_{DD'}, M_{DT'}, M_{TT'},$ and $M_{TD'}$, respectively.
Since the breaking $G_{321}\tm {G_{321}}'\to G_{SM}\tm SU(3)'$ is realized by the VEV of the fragment $\Phi_{DD'}$
at scale $M_I$, we take $M_{DD'}\simeq M_I$. The state $\Phi_3$, participating in Eq. (\ref{triplet-mixing}), emerges from this $\Phi_{DD'}$ fragment.
The remaining three states under $G_{321}\tm SU(3)'$ transform as
\beq
G_{321}\tm SU(3)'~:~~
\Phi_{DT'}\sim \l 1, 2, -\fr{5}{\sq{60}}, \bar 3'\r ,~~
~ \Phi_{TT'} \sim \l 3, 1, 0, \bar 3'\r ,~~
\Phi_{TD'}\sim \l 3, 2, \fr{5}{\sq{60}}, 1\r .
\la{Phi-frag-transSM}
\eeq
The mass squares of these fields are given by
\beq
M_{DT'}^2=5\lam_{2\Si \Phi}v_{\Si'}^2~,~~~M_{TT'}^2=5\lam_{2\Si \Phi}(v_{\Si}^2+v_{\Si'}^2)\!-\!2\lam_{2\Phi}v_{\Phi }^2~,~~~
M_{TD'}^2=5\lam_{2\Si \Phi}v_{\Si }^2~.
\la{Phi-fragm-mass}
\eeq

With  the VEVs toward  the directions given in Eqs. (\ref{Si-VEV}), (\ref{Si1-VEV}), and (\ref{vev-conf}), and with the extremum conditions of Eq. (\ref{min-conditions}),
 the potential's minimum is achieved with
$$
30\lam_1+7\lam_2>0~,~~~\lam_2>0~,~~~\lam >0~,
$$
\beq
10\lam_{1\Si \Phi }-3\lam_{2\Si \Phi }>0~,~~~\lam_{2\Si \Phi }>0~,~~~2\lam_{1\Phi}+\lam_{2\Phi}>0~,~~~\lam_{2\Phi}>0~.
\la{stab-min}
\eeq

As far as the states $H$ and $H'$ are concerned, they are split as $H\to (D_H, T_H)$ and $H'\to (D_{H'}, T_{H'})$,
where $D_H, D_{H'}$ are doublets, while $T_H$ and $T_{H'}$ are $SU(3)_c$ and $SU(3)'$ triplets, respectively.
Mass squares of these triplets are
$$
M_{T_H}^2=M_{H}^2-4h_1v_{\Si }^2+30(h_2v_{\Si }^2+\tl{h}v_{\Si' }^2)+2\lam_{1H\Phi}v_{\Phi}^2/\sqrt{25}~,
$$
\beq
M_{T_{H'}}^2=M_{H}^2-4h_1v_{\Si' }^2+30(h_2v_{\Si' }^2+\tl{h}v_{\Si }^2)+2\lam_{1H\Phi}v_{\Phi}^2/\sqrt{25}~.
\la{T-masses}
\eeq
The states $D_H$ and $D_{H'}$, under $G_{SM}$, both have quantum numbers of the SM Higgs doublet. They mix by the VEV $\lan \Phi \ran$,
and the mass squared matrix is given by
\beq
 \l D_{H}^\dag, ~ D_{H'}^\dag\r \!
\left( \!
  \begin{array}{cc}
    M_{T_H}^2\!\!-\!5h_1v_{\Si }^2\!+\!\lam_{2H\Phi}v_{\Phi}^2/\sqrt{10} & \mu v_{\Phi} \\
   \mu v_{\Phi} & M_{T_{H'}}^2\!\!-\!5h_1v_{\Si' }^2\!+\!\lam_{2H\Phi}v_{\Phi}^2/\sqrt{10} \\
  \end{array}
\! \right)\!\!
\l \!\!\!
  \begin{array}{c}
    D_{H} \\
    D_{H'} \\
  \end{array}\!\!\!\r .
\la{HD-mixing}
\eeq
By diagonalization of  (\ref{HD-mixing}), we get two physical states $h$ and $D'$:
$$
h=\cos \te_hD_H+\sin \te_hD_{H'}~,~~~D'=-\sin \te_hD_H+\cos \te_hD_{H'}~,
$$
\beq
\tan 2\te_h=\fr{2\mu v_{\Phi}}{M_{T_H}^2-M_{T_{H'}}^2-5h_1(v_{\Si }^2-v_{\Si' }^2)}~.
\la{hD-te-h}
\eeq
We identify $h$ with the SM Higgs doublet and set its mass square (by fine-tuning) $M_h^2\sim 100~{\rm GeV}^2$.  We assume the second doublet $D'$ to be
 heavy $M_{D'}^2\gg |M_h|^2$. For the mixing angle $\te_h$, we also assume  $\te_h\ll 1$. Therefore, according to Eq. (\ref{hD-te-h}), the SM Higgs mainly
resides in $D_H$ (of the $H$-plet), while $D_{H'}$ (i.e., $H'$) includes a light SM doublet with very suppressed weight.

The radiative corrections will affect obtained expressions for the masses and VEVs. However, there are enough parameters
involved, and one can always get considered symmetry breaking pattern and desirable spectrum.
Achieving these will require some fine-tunings.
Without addressing here the hierarchy problem and naturalness issues, we will proceed to study various properties and the phenomenology
of the considered scenario.

\section{Fermion sector}
\la{ferm}

\subsection{$D_2$ symmetry  $\grave{\rm \bf a}$ la $P$ parity}

We introduce three families of $(\Psi, F)$ and three families of $(\Psi', F')$,
\beq
3\tm \left [ \Psi (10, 1)+F(\bar 5, 1)\right ]~,\hs{0.6cm} ~~
3\tm \left [ \Psi'(1, \bar{10})+F'(1,  5)\right ]~,
\la{matter}
\eeq
where in brackets the transformation properties under $SU(5)\tm SU(5)'$ gauge symmetry are indicated. Here,
each fermionic state is a two-component Weyl spinor, in $(\fr{1}{2},0)$ representation of the Lorentz group.
The action of $D_2$ parity on these fields is determined as
\beq
D_2~:~~~~~~\Psi \stackrel{\rarr}{_\larr} \ov{\Psi'}\equiv (\Psi')^\dag~,~~~
\F \stackrel{\rarr}{_\larr} \ov{F'}\equiv (F')^\dag~.
\la{matter-D2}
\eeq
It is easy to verify that, with transformations in Eqs. (\ref{matter-D2}) and (\ref{D5gauge}), the kinetic part of
the Lagrangian ${\cal L}_{kin}(\Psi,F,\Psi',F')$ is invariant.\footnote{The $D_2$ transformation of Eq. (\ref{matter-D2}) resembles
usual $P$ parity, acting between the electron and positron, within QED. Unlike the QED,   the states $(\Psi, F)$ and $(\Psi ',F')$
 transform under different gauge groups.}

We can easily   write down invariant Yukawa Lagrangian
\beq
{\cal L}_Y+{\cal L}_{Y'}+{\cal L}_Y^{mix}
\la{L-Y-tot}
\eeq
with
\beq
{\cal L}_Y=\sum_{n=0}C_{\Psi\Psi}^{(n)}\l \fr{\Si }{M_*}\r^n\Psi \Psi H+
\sum_{n=0}C_{\Psi F}^{(n)}\l \fr{\Si }{M_*}\r^n\Psi \F H^\dag +{\rm h.c.}
\la{L-Y}
\eeq
\beq
{\cal L}_{Y'}=\sum_{n=0}C_{\Psi\Psi}^{(n)*}\l \fr{\Si' }{M_*}\r^n\Psi' \Psi' {H'}^\dag +
\sum_{n=0}C_{\Psi F}^{(n)*}\l \fr{\Si' }{M_*}\r^n\Psi' \F' H' +{\rm h.c.}
\la{L-Y1}
\eeq
\beq
{\cal L}_Y^{mix}=\lam_{FF'}F\Phi F'+\lam_{FF'} \ov{F'}\Phi^\dag \ov{F}+
\fr{\lam_{\Psi \Psi'}}{M}\Psi (\Phi^\dag )^2\Psi' +\fr{\lam_{\Psi \Psi'}}{M}\ov{\Psi'}\Phi^2\ov{\Psi }~,
\la{L-Ymix}
\eeq
where $M_*, M$ are some cutoff scales.
The coupling matrices  $\lam_{FF'}$ and $\lam_{\Psi \Psi'}$ are Hermitian due to the $D_2$ symmetry.
The last two higher-order operators in Eq. (\ref{L-Ymix}), important for phenomenology, can be generated by integrating out some heavy
states with mass at or above the GUT scale. For instance, with the scalar state $\Om $ in $(\bar{10},10)$ representation of
$SU(5)\tm SU(5)'$ and $D_2$ parity, $\Om \stackrel{\rarr}{_\larr} \Om^\dag $, the relevant terms (of fundamental Lagrangian) will be
$\lam_{\Psi \Psi'}\Om \Psi \Psi' +\lam_{\Psi \Psi'} \Om^\dag \ov{\Psi'}\!\! \cdot \!\ov{\Psi }+
\bar M_{\Om}(\Om \Phi^2+\Om^\dag (\Phi^\dag )^2)+M_{\Om}^2\Om^\dag \Om $.
With these couplings, one can easily verify that integration of $\Om $ generates the last two operators of Eq. (\ref{L-Ymix})
(with $M\approx M_{\Om}^2/\bar M_{\Om}$). Since the $\Om$ is rather heavy, its only low-energy implication can be the emergence of these effective operators.
Thus, in our further studies, we will proceed with the consideration of Yukawa couplings given in Eqs. (\ref{L-Y})-(\ref{L-Ymix}).

With obvious identifications, let us adopt the following notations for the components from $\Psi, F$ and $\Psi', F'$ states:
$$
\Psi =\{ q, u^c, e^c\}~,~~~~F=\{ l, d^c\}~,
$$
\beq
\Psi' =\{ \hat q, \hat u^c, \hat e^c\}~,~~~~F'=\{ \hat l, \hat d^c\}~.
\la{matter-comp}
\eeq
Substituting in Eqs. (\ref{L-Y})-(\ref{L-Ymix}) the VEVs  $\lan \Si\ran , \lan \Si'\ran $, and $\lan \Phi \ran $,  the relevant couplings we obtain are
$$
{\cal L}_Y\to q^TY_Uu^ch+q^TY_Dd^ch^\dag +e^{cT}Y_{e^cl}lh^\dag +
$$
\beq
(C_{qq}qq+C_{u^ce^c}u^ce^c)T_H+(C_{ql}ql+C_{u^cd^c}u^cd^c)T_H^\dag +
{\rm h.c.}
\la{from-LY}
\eeq
\beq
{\cal L}_{Y'}\to C_{\Psi\Psi}^{(0)*}(\fr{1}{2}\hat q\hat q+\hat u^c\hat e^c) T_{H'}^\dag +
 C_{\Psi F}^{(0)*}(\hat q\hat l+\hat u^c \hat d^c)T_{H'}
 +{\rm h.c.}+\cdots
\la{from-LY1}
\eeq
\beq
{\cal L}_Y^{mix}\to \hat l^TM_{\hat ll}l+e^{cT}M_{e^c\hat e^c}\hat e^c+{\rm h.c.}~.
\la{from-LYmix}
\eeq
 In Eq. (\ref{from-LY1}) we have dropped out the couplings with the Higgs doublet because, as we have assumed, $D_{H'}$ includes the SM Higgs doublet
 with very suppressed weight. Also, we have ignored powers of $\lan \Si'\ran/M_*$ in comparison with $\lan \Si\ran/M_*$'s exponents.
As we will see, the couplings of $h$ in (\ref{from-LY}) and terms shown in Eqs. (\ref{from-LY1}) and (\ref{from-LYmix}) are responsible for
fermion masses and mixings and lead to realistic phenomenology.

\subsection{Fermion masses and mixings: Composite leptons}
\la{fermion-mass}

Let us first indicate transformation properties of all matter states, given in Eq. (\ref{matter-comp}), under the unbroken
 $G_{SM}\tm SU(3)'=SU(3)_c\tm SU(2)_w\tm U(1)_Y\tm SU(3)'$ gauge symmetry. Fragments from $\Psi, F$ transform as
$$
q\sim (3,2, -\fr{1}{\sq{60}}, 1)~,~~~u^c\sim (\bar 3, 1, \fr{4}{\sq{60}}, 1)~,~~~e^c\sim (1,1,-\fr{6}{\sq{60}}, 1)
$$
\beq
l\sim (1, 2, \fr{3}{\sq{60}}, 1)~,~~~~~~d^c\sim (\bar 3, 1, -\fr{2}{\sq{60}}, 1)~,
\la{SM-matter-trans}
\eeq
while the states from $\Psi', F'$ have the following transformation properties:
$$
\hat q\sim (1,2, \fr{1}{\sq{60}}, \bar 3')~,~~~\hat u^c\sim (1, 1, -\fr{4}{\sq{60}}, 3')~,~~~
\hat e^c\sim (1,1,\fr{6}{\sq{60}}, 1)
$$
\beq
\hat l\sim (1, 2, -\fr{3}{\sq{60}}, 1)~,~~~~~~\hat d^c\sim (1, 1, \fr{2}{\sq{60}}, 3')~.
\la{hat-matter-trans}
\eeq
In transformation properties of Eq. (\ref{hat-matter-trans}), by primes we have indicated triplets and antitriplets of $SU(3)'$.
As we see, transformation properties of quark states in Eq. (\ref{SM-matter-trans}) coincide with those of the SM. Therefore,
for quark masses and CKM mixings,  the first two couplings of Eq. (\ref{from-LY}) are relevant.
Since in $Y_{U,D}$ and $Y_{e^cl}$ contribute also higher-dimensional operators, the $Y_U$ is not symmetric and
$Y_D\neq Y_{e^cl}$. Thus, quark Yukawa matrices can be diagonalized by biunitary transformations
\beq
L_u^\dag Y_UR_u=Y_U^{\rm Diag}~,~~~~L_d^\dag Y_DR_d=Y_D^{\rm Diag}~.
\la{YUD-diag}
\eeq
With these, the CKM matrix (in standard parametrization) is
$$
V_{CKM}=P_1L_u^TL_d^*P_2
$$
\beq
{\rm with}~~~P_1\!=\!{\rm Diag}\!\l e^{i\om_1}, ~e^{i\om_2},~e^{i\om_3}\r ~,~~~~~~~
P_2\!=\!{\rm Diag}\!\l e^{i\rho_1}, ~e^{i\rho_2},~1\r ~.
\la{ckm}
\eeq

\vspace{0.4cm}
{\bf Composite leptons}
\vspace{0.15cm}

Turning to the lepton sector, we note that  $\hat l$  and $\hat e^c$ have opposite/conjugate transformation
properties with respect to  $l$ and  $e^c$, respectively. From couplings in Eq. (\ref{from-LYmix}), we see that these vectorlike states
acquire masses $M_{\hat ll}$ and $M_{e^c\hat e^c}$ and decouple . However,  within this scenario, composite leptons
emerge. The $SU(3)'$ becomes strongly coupled and confines at scale $\La' \sim $~TeV (for details, see Sec. \ref{unif}).
 Because of confinement, $SU(3)'$ singlet composite states -  baryons ($B'$) and/or mesons ($M'$)  - can emerge.
The elegant idea
of fermion emergence through the strong dynamics as bound states of more fundamental constituents, was suggested and developed
in Refs. \cite{Pati:1975md}-\cite{review-comp}. Within our scenario,
this idea finds an interesting realization for the lepton states.
Formation of composite fermions should satisfy 't Hooft anomaly matching conditions\footnote{In case
the chiral symmetry remains unbroken (at least partially) at the composite level. The models avoiding anomaly conditions
were suggested in Ref. \cite{comp-no-anom-cond}.}
 \cite{hooft}. These give a severe constraint on
building models with composite fermions \cite{Dimopoulos:1980hn,Nilles:1981bx,Greenberg:1980ri},
\cite{Buchmuller:1983iu,Chkareuli:1982rn,review-comp}.

Let us focus on the sector of (three-family) $\hat q, \hat u^c$ and $\hat d^c$ states, which have $SU(3)'$ strong interactions. Ignoring
local EW  and Yukawa interactions, the Lagrangian of these states possesses global $G_f^{(6)}=SU(6)_L\tm SU(6)_R\tm U(1)_{B'}$   chiral symmetry.
Under the $SU(6)_L$, three families of $\hat q=(\hat u, \hat d)$ transform as sextet $6_L$, while three families of
$(\hat u^c, \hat d^c)\equiv \hat q^c$ form sextet $6_R$ of $SU(6)_R$. The $U(1)_{B'}$ ($B'$) charges of $\hat q$ and $\hat q^c$ are,
respectively, $1/3$ and $-1/3$. Thus, transformation properties of these states under
\beq
G_f^{(6)}=SU(6)_L\tm SU(6)_R\tm U(1)_{B'}
\la{Gf6}
\eeq
chiral symmetry are
\beq
\hat q_{\al } =(\hat u, \hat d)_{\al }\sim (6_L, 1, \fr{1}{3})~,~~~~~
\hat q^c_{\al }=(\hat u^c, \hat d^c)_{\al } \sim (1, 6_R, -\fr{1}{3})~,
\la{Gf-transf}
\eeq
where $\al =1,2,3$ is the family index.
 Because of the strong $SU(3)'$ attractive force,  condensates that will break the $G_f^{(6)}$ chiral symmetry can form. The breaking
can occur by several steps, and at each step the formed composite states should satisfy anomaly matching conditions.

In Appedix \ref{comp-anom}, we give a detailed account of these issues and demonstrate that within our scenario
 three families of  $l_0, e^c_0, \nu^c_0$ composite states,
\beq
(\hat q \hat q)\hat q \sim l_{0\al }=\!\!\l \!\!
                                       \begin{array}{c}
                                         \nu_0 \\
                                         e_0 \\
                                       \end{array}
                                     \!\!\r_{\al }
~,~~~~
(\hat q^c \hat q^c)\hat q^c =\l (\hat u^c \hat d^c)\hat d^c,~(\hat u^c \hat d^c)\hat u^c\r
\sim l^c_{0\al}\equiv (\nu^c_0, ~e^c_0)_{\al}~,~~~~
\al =1,2,3
\la{comp-st}
\eeq
emerge. In Eq. (\ref{comp-st}), for combinations $(\hat q \hat q)\hat q$ and $(\hat q^c \hat q^c)\hat q^c$, the spin-1/2 states are assumed with
suppressed gauge and/or flavor indices. For instance, under
$(\hat q \hat q)\hat q$  we mean
$\ep^{a'b'c'}\!\ep_{ij}(\hat q_{a'i} \hat q_{b'j})\hat q_{c'k}$, where $a', b', c'=1,2,3$ are $SU(3)'$ indices and
$i,j,k=1,2$ stand for $SU(2)_w$ (or $SU(2)_L$) indices.
Similar  applies to the combination $(\hat q^c \hat q^c)\hat q^c$.
Thus,  $(\hat q \hat q)\hat q$ and $(\hat q^c \hat q^c)\hat q^c$ are singlets of $SU(3)'$.
From these, taking into account Eqs. (\ref{hat-matter-trans}) and (\ref{comp-st}), it is easy to verify that the quantum numbers of
composite states under SM gauge group
$G_{SM}=SU(3)_c\tm SU(2)_w\tm U(1)_Y$ are
\beq
G_{SM}~:~~~l_0\sim (1, 2, \fr{3}{\sq{60}})~,~~~e^c_0\sim (1, 1, -\fr{6}{\sq{60}})~,~~~\nu^c_0\sim (1, 1, 0)~.
\la{comp-transl}
\eeq
As we see, along with SM leptons ($l_0$ and $e^c_0$), we get three families of composite SM singlets fermions - $\nu^c_0$.
The latter can be treated as composite
right-handed/sterile neutrinos in the spirit of Ref. \cite{ArkaniHamed:1998pf}.
Note that, with this composition, as was expected, the gauge anomalies also vanish (together with the chiral anomaly matching;
for details, see Appendix \ref{comp-anom}).
Interestingly, the $SU(3)'$ [originating from $SU(5)'$] triplet and antitriplets
$\hat u^c, \hat d^c$ and $\hat q$
play the role of "preon" constituents for the bound-state leptons and right-handed/sterile neutrinos.
Moreover, in our scheme the lepton number $L$ is related to the $U(1)_{B'}$
charge as $L=3B'$. Therefore,  "primed baryon number" $B'$ [of the $SU(5)'$] is the origin of the lepton number.

\vspace{0.4cm}
{\bf Charged lepton masses}
\vspace{0.15cm}

Now, we turn to the masses of the charged leptons, which are composite within our scenario. As it turns out,
their mass generation does not require additional extension. It happens via integration of the states that are present in the
model. As we see from Eq. (\ref{from-LY1}), the $SU(5)'$ matter  couples with the $SU(3)'$ triplet scalar $T_{H'}$ with mass $M_{T_{H'}}$.
Relevant 4-fermion operators, emerging from the couplings of Eq. (\ref{from-LY1}) and by integration of  $T_{H'}$, are
\beq
{\cal L}_{Y'}^{eff}=\fr{C_{\Psi\Psi}^{(0)*}C_{\Psi F}^{(0)*}}{M_{T_{H'}}^2}
\left [ \fr{1}{2}(\hat q\hat q)(\hat q\hat l)+(\hat u^c\hat e^c)(\hat u^c\hat d^c)\right ] +{\rm h.c.}
\la{L-Y1eff}
\eeq
As we see, here appear the combinations $(\hat q \hat q)\hat q$ and $(\hat u^c\hat d^c)\hat u^c$, which according
to Eq. (\ref{comp-st}) form composite charged lepton states. We will use the parametrizations
\beq
\fr{1}{2}(\hat q_{\al }\hat q_{\bt })\hat q_{\ga }={\La'}^3c_{\al \bt\ga\de}l_{0\de}~,~~~~~
(\hat u^c_{\al }\hat d^c_{\bt })\hat u^c_{\ga }={\La'}^3\bar c_{\al \bt\ga\de}e^c_{0\de}~
\la{comp-lept}
\eeq
where Greek indices denote family indices and  $c, \bar c$ are dimensionless couplings - four index  tensors in a family space. The  $(l_0, e^c_0)_{\de }$ denote three families of composite leptons.
Using Eq. (\ref{comp-lept}) in Eq. (\ref{L-Y1eff}), we obtain
$$
{\cal L}_{Y'}^{eff}\to \hat l\hat{\mu}l_0+e^c_0\tl{\mu}\hat e^c+{\rm h.c.}
$$
\beq
{\rm with}~~~~
\hat{\mu}_{\de'\de}\equiv\fr{{\La'}^3}{M_{T_{H'}}^2}(C_{\Psi\Psi}^{(0)*})_{\al\bt}(C_{\Psi F}^{(0)*})_{\ga\de'}
c_{\al\bt\ga\de}~,
~~~~~
\tl{\mu}_{\de\de'}\equiv\fr{{\La'}^3}{M_{T_{H'}}^2}(C_{\Psi\Psi}^{(0)*})_{\ga\de'}(C_{\Psi F}^{(0)*})_{\al\bt}
\bar c_{\al\bt\ga\de}~.
\la{L-Y1eff-1}
\eeq
\begin{figure}[t]
\begin{center}
\hs{-1cm}
\resizebox{0.9\textwidth}{!}{
 \hs{3cm} \vs{1cm}\includegraphics{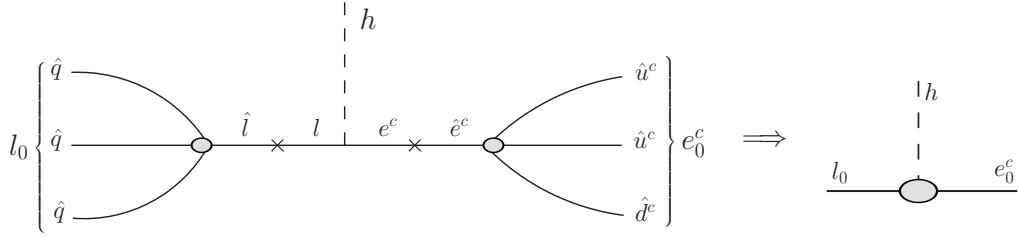}
}
\vs{0.7cm}
\caption{Diagram responsible for the generation of the charged lepton effective Yukawa matrix.}
\label{fig1}       
\end{center}
\end{figure}
At the next stage, we integrate out the vectorlike states $\hat l, l$ and $e^c, \hat e^c$,
which, respectively, receive masses $M_{\hat ll}$ and $M_{e^c\hat e^c}$  through the coupling in
Eq. (\ref{from-LYmix}).
Integrating out these heavy states, from  Eqs. (\ref{from-LYmix}) and (\ref{L-Y1eff-1}), we get
\beq
l\simeq -\fr{1}{M_{\hat ll}}\hat{\mu}l_0~,~~~~e^{cT}\simeq -e^{cT}_0\tl{\mu}\fr{1}{M_{e^c\hat e^c}}~.
\la{int-lec}
\eeq
Substituting these in the $e^{cT}Y_{e^cl}lh^\dag$ coupling of Eq. (\ref{from-LY}), we see that the effective  Yukawa couplings for the leptons are generated:
\beq
l_0^TY_Ee^c_0h^\dagger +{\rm h.c.}~~~~~~
{\rm with}~~~~ Y_E^T\simeq \tl{\mu}\fr{1}{M_{e^c\hat e^c}}Y_{e^cl}\fr{1}{M_{\hat ll}}\hat{\mu}~.
\la{YE}
\eeq
The diagram corresponding to the generation of this effective Yukawa operator is shown in Fig. \ref{fig1}.
This mechanism is novel and differs from those suggested earlier for the mass generation of composite fermions
\cite{review-comp}. From the observed values of the Yukawa couplings, we have
$|{\rm Det}Y_E|=\lam_e\lam_{\mu}\lam_{\tau}\approx 1.8\cdot 10^{-11}$. On the other hand, natural values of the eigenvalues of
$Y_{e^cl}$ can be$\sim 0.1$. Thus,  $|{\rm Det}Y_{e^cl}|\sim 10^{-3}$. From these and the expression given in Eq. (\ref{YE}),
we obtain
\beq
|{\rm Det} (\tl{\mu}\fr{1}{M_{e^c\hat e^c}})|\cdot
|{\rm Det}(\fr{1}{M_{\hat ll}}\hat{\mu})|\sim 10^{-8}~~,
\la{constr-matr}
\eeq
the constraint that should be satisfied by two matrices $\tl{\mu}\fr{1}{M_{e^c\hat e^c}}$ and $\fr{1}{M_{\hat ll}}\hat{\mu}$.

\vspace{0.4cm}
{\bf Neutrino masses}
\vspace{0.15cm}

Now, we discuss the neutrino mass generation. To accommodate the neutrino data \cite{nu-data},
one can use SM singlet fermionic states in order to generate either Majorana- or
Dirac-type masses for the neutrinos. Within our model,
among the composite fermions, we have SM singlets $\nu^c_0$ [see Eqs. (\ref{comp-st}) and (\ref{comp-transl})]. Here, we stick to the possibility of the Dirac-type neutrino masses, which can be naturally suppressed \cite{ArkaniHamed:1998pf}. Because of compositeness, there is no direct Dirac couplings $Y_{\nu }$ of $\nu^c_0$'s with lepton doublets $l_0$. Similar to the charged lepton Yukawa couplings, we need to generate $Y_{\nu }$. For this purpose, we introduce the $SU(5)\tm SU(5)'$ singlet
(two-component) fermionic
states $N$.\footnote{The number of $N$ states is not limited, but for simplicity we can assume that they are not more than 3.}
Assigning the $D_2$ parity transformations $N \stackrel{\rarr}{_\larr} \ov{N}$ and taking into account Eqs. (\ref{sc-D2-tr}) and
(\ref{matter-D2}),
relevant couplings, allowed by  $SU(5)\tm SU(5)'\tm D_2$ symmetry, will be
\beq
{\cal L}_N=C_{FN}FNH+C_{FN}^*F'N{H'}^\dag -\fr{1}{2}N^TM_NN+{\rm h.c.}~~~~
{\rm with}~~~~M_N=M_N^*~.
\la{L-N}
\eeq
These give the following interaction terms:
\beq
{\cal L}_N\to C_{FN}lNh+C_{FN}^*\hat d^cNT_{H'}^\dag -\fr{1}{2}N^TM_NN+{\rm h.c.}
\la{from-L-N}
\eeq
From these and Eq. (\ref{from-LY1}), integration of $T_{H'}$ state gives  additional affective four-fermion operator
\beq
\fr{C_{\Psi F}^{(0)*}C_{FN}^*}{M_{T_{H'}}^2}(\hat u^c \hat d^c)(\hat d^c N)+{\rm h.c.}
\la{eff-N}
\eeq
By the parametrization
\beq
(\hat u^c_{\al }\hat d^c_{\bt })\hat d^c_{\ga }={\La'}^3\tl c_{\al \bt\ga\de}\nu^c_{0\de}~,
\la{comp-nuc}
\eeq
operators in Eq. (\ref{eff-N}) are given by
\beq
{\cal L}_{N\nu^c}^{eff}=N\mu_{\nu}\nu^c_0+{\rm h.c.}~~~~
{\rm with}~~~
(\mu_{\nu})_{\de'\de}\equiv \fr{{\La'}^3}{M_{T_{H'}}^2}
(C_{\Psi F}^{(0)*})_{\al\bt}(C_{FN}^*)_{\ga\de'}\tl c_{\al \bt\ga\de}~.
\la{eff-Nnuc}
\eeq
Subsequent integration of $N$ states,  from Eq. (\ref{eff-Nnuc}) and the last term of Eq. (\ref{from-L-N}) gives
\beq
N\simeq \fr{1}{M_N}\mu_{\nu}\nu^c_0~.
\la{int-N}
\eeq

Substituting this, and the expression of $l$ from Eq. (\ref{int-lec}), in the first term of Eq. (\ref{from-L-N}), we arrive at
\beq
l_0^TY_{\nu}\nu^c_0h +{\rm h.c.}~~~~~
{\rm with}~~~~ Y_{\nu}\simeq -\hat{\mu}^T\fr{1}{M_{\hat ll}^T}C_{FN}\fr{1}{M_N}\mu_{\nu}~.
\la{Ynu}
\eeq

\begin{figure}[t]
\begin{center}
\hs{-1cm}
\resizebox{0.9\textwidth}{!}{
 \hs{3cm} \vs{1cm}\includegraphics{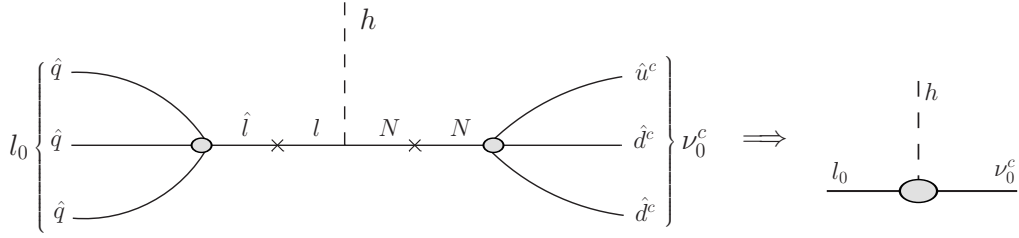}
}
\vs{0.7cm}
\caption{Diagram responsible for the generation of the effective Dirac Yukawa matrix for the neutrinos.}
\label{fig2}       
\end{center}
\end{figure}

The relevant diagram generating this effective Dirac Yukawa couplings is given in Fig. \ref{fig2}.
With  $\fr{1}{M_{\hat ll}}\hat{\mu}\sim 10^{-2}$ and $C_{FN}\sim M_N\sim \fr{1}{M_N}\mu_{\nu}\sim 10^{-5}$,  we can get the Dirac neutrino
mass $M_{\nu}^D=Y_{\nu}\lan h^{(0)}\ran \sim 0.1$~eV, which is the right scale to explain neutrino anomalies.
Note that using Eq. (\ref{int-N}) in the last term of Eq. (\ref{from-L-N}) we also obtain the term $-\fr{1}{2}{\nu^c_0}^TM_{\nu^c}\nu^c_0$ with $M_{\nu^c}\simeq \mu_{\nu}^T\fr{1}{M_N}\mu_{\nu}$. By proper
 selection of the couplings $C_{FN}$ and eigenvalues of $M_N$, the $M_{\nu^c}$ can be strongly suppressed. In this case, the neutrinos
 will be (quasi)Dirac. However, it is possible that some of the species of light neutrinos to be (quasi)Dirac and some of them Majoranas.
 Detailed studies of such scenarios and their compatibilities with current experiments \cite{Abazajian:2012ys}
 are beyond the scope of this paper.

\section{Gauge coupling unification}
\la{unif}

In this section we will study the gauge coupling unification within our model. We show that the
symmetry breaking pattern  gives the possibility for successful unification.\footnote{Possibilities of gauge coupling unification, with the intermediate symmetry breaking pattern and  without invoking low-scale supersymmetry,  have been studied in Ref. \cite{interm-unif-gut}.}
 As it turns out, the
 $SU(3)'$ gauge interaction becomes strongly coupled at scale  $\La'$($\sim $few TeV).
Thus,  below this scale, $SU(3)'$ confines, and all states (including composite ones) are $SU(3)'$ singlets.
 Therefore, with the masses $M_{\hat ll}^{(\al)}$ and $M_{e^c\hat e^c}^{(\al)}$ ($\al =1,2,3$) of vectorlike
states $l, \hat l$ and $e^c, \hat e^c$ being above the scale $\La'$,
 in the energy interval $\mu =M_Z-\La'$, the states are just those of SM (plus possibly
right-handed/sterile neutrinos having no impact on gauge coupling running), and corresponding one-loop $\bt$-function coefficients
are $(b_Y, b_w, b_c)=\l \fr{41}{10}, -\fr{19}{6}, -7\r $.
Since $\La'$ is the characteristic scale of the strong dynamics, it is clear that pseudo-Goldstone and composite states (besides
SM leptons) emerging through chiral symmetry breaking and strong dynamics, can have masses below  $\La'$ (in a certain range). Instead
investigating their spectrum and dealing with corresponding threshold effects, we parametrize all these as a single effective $\La'$
scale, below which theory is the SM. This phenomenological simplification allows us to proceed with RG analysis. Note, however, that even with taking those kinds of
thresholds into account should not harm the success of coupling unification with the price of proper adjustment of the mass scales
 (given in Table \ref{t:tab1} and discussed later on).

In the energy interval $\La' - M_I$, we have the symmetry $SU(3)_c\tm SU(2)_w\tm U(1)_Y\tm SU(3)'$, and
 $SU(3)'$ nonsinglet states (i.e., $\hat q, \hat u^c, \hat d^c$, $T_{H'}$, etc.) must be taken into account.
As was noted in Sec. \ref{55model}, we consider hierarchical breaking:   $M_I\ll {M_I}'\ll M_G$ [see Eqs. (\ref{MI-MI1}) and (\ref{VEV-hier})].
 This choice
allows us to have successful unification with confining scale $\La'\sim $ few TeV.\footnote{One can have unification with $\lan \Si'\ran =0$,
(i.e., $M_I={M_I}'$) and with a modified spectrum. However, with such a choice the value of $\La' $
comes out rather large ($\stackrel{>}{_\sim }10^5$~GeV). This would also imply the breaking of EW symmetry at a high scale and thus should be discarded from the phenomenological viewpoint. More discussion about this issue is given in Sec. \ref{impl}.}
Thus, between the scales $M_I$ and ${M_I}'$, the symmetry is $G_{321}\tm {G_{321}}'$ [see Eqs.  (\ref{stage1}) and (\ref{stage2})],
and states should be decomposed under these groups [see, for instance, Eqs. (\ref{Phi55-dec}) and (\ref{Phi-frag-trans})].
Since the breaking $G_{321}\tm {G_{321}}'\to G_{SM}\tm SU(3)'$ is realized by the VEV of the fragment $\Phi_{DD'}$
at scale $M_I$, we take $M_{DD'}\simeq M_I$. The remaining three masses, of the fragments coming from $\Phi $,  can be in a range $\La' - M_G$.
Giving more detailed account to these issues in Appendix \ref{rg}, below we sketch the main details.

Above  the scale $M_I$, all matter states should be included in the RG.
Above the scale ${M_I}'$ we have the $SU(5)'$ symmetry, and
 the fragments $\Phi_{DD'}, \Phi_{DT'}$ form the unified $(2,\bar 5)$-plet of $G_{321}\tm SU(5)'$:
 $(\Phi_{DD'}, \Phi_{DT'})\subset \Phi_{D\bar 5'}$, while  $\Phi_{TT'}$ and $\Phi_{TD'}$ states unify in $(3,\bar 5)$-plet:
 $(\Phi_{TT'}, \Phi_{TD'})\subset \Phi_{T\bar 5'}$.
 These states, together with the $\Si'$-plet, should be included in the RG above the scale ${M_I}'$.

According to Eq. (\ref{g-match-MI}), at scale $M_I$, for the EW gauge couplings, we have the boundary conditions
\beq
\al_Y^{-1}(M_I)=\al_1^{-1}(M_I)+\al_{1'}^{-1}(M_I)~,~~~\al_w^{-1}(M_I)=\al_2^{-1}(M_I)+\al_{2'}^{-1}(M_I)~.
\la{bound-Yw}
\eeq
The couplings of ${G_{321}}'$ gauge interactions unify and form single  $SU(5)'$ coupling at scale ${M_I}'$ :
\beq
\al_{1'}({M_I}')=\al_{2'}({M_I}')=\al_{3'}({M_I}')= \al_{5'}({M_I}')~.
\la{MI1-unif}
\eeq
Finally, at the GUT scale $M_G$, the coupling of  $G_{321}$ and $SU(5)'$ unifies:
\beq
 \al_1(M_G)=\al_2(M_G)=\al_3(M_G)=\al_{5'}(M_G)\equiv \al_G ~.
\la{unif-MG}
\eeq
%
%
%
\begin{table}
\caption{Particle spectroscopy.
 }

\label{t:tab1} $$\begin{array}{|c|c||c|c||c|c||c|c||c|c|}

\hline
\vs{-0.3cm}
 &  &  &  &  &  &  &  && \\

\vs{-0.4cm}

M_a& \hs{0.3mm}{\rm GeV}\hs{0.3mm}& \hs{0.3mm}M_a\hs{0.3mm}&\hs{0.3mm}{\rm GeV}\hs{0.3mm}& \hs{0.5mm}M_a \hs{0.5mm}&
\hs{0.5mm}{\rm GeV}\hs{0.5mm} &\hs{0.5mm}M_a\hs{0.5mm} &\hs{-0.5mm}{\rm GeV} \hs{-0.5mm}&\hs{-0.5mm}M_a\hs{-0.5mm}
  & \hs{-0.5mm}{\rm GeV}\hs{-0.5mm}\\

&  &  &  &  &  &  &  &  & \\

\hline
\hline

\vs{-0.3cm}
 &  &  &  &  &  &  &  & & \\

\vs{-0.3cm}
\hs{-0.5mm}M_{\hat ll}^{(1)}\hs{-0.5mm}& 7.54\cdot 10^4 & M_{e^c\hat{e}^c}^{(2)}& 7.54\cdot 10^4  & M_{D'} & 4.16\cdot 10^6 &
 M_{TD'} &3.92\cdot 10^6 &
M_{X'} & 2.08\cdot 10^6  \\

&  &  &  &  &  &  &  & & \\

\hline
\vs{-0.3cm}
 &  &  &  &  &  &  &  & & \\

\vs{-0.3cm}
\hs{-0.5mm}M_{\hat ll}^{(2)}\hs{-0.5mm}& 7.54\cdot 10^4 & M_{e^c\hat{e}^c}^{(3)} & 1.2\cdot 10^5 & M_{TT'} & 1874.7 & M_{\Si'_{8'}} &9277
&M_{T_H} &5\cdot 10^{11}\\

&  &  &  &  &  &  &  & & \\

\hline
\vs{-0.3cm}
 &  &  &  &  &  &  &  & & \\

\vs{-0.3cm}
\hs{-0.5mm}M_{\hat ll}^{(3)}\hs{-0.5mm}& 1.2\cdot 10^5 & \La'& 1851  & M_{DD'} & 8.25\cdot 10^4 & M_{\Si'_{3'}} &2M_{\Si'_{8'}} &M_{X} &4.95 \cdot 10^{11}   \\

&  &  &  &  &  &  &  & &\\

\hline
\vs{-0.3cm}
 &  &  &  &  &  &  &  & & \\

\vs{-0.3cm}
\hs{-0.5mm}M_{e^c\hat{e}^c}^{(1)}\hs{-0.5mm}& 7.54\cdot 10^4 & M_{T_{H'}} & 1851  & M_{DT'} & 8250 & M_{\Si'_{1'}} &4.16\cdot 10^6
&M_{\Si} &5 \cdot 10^{11}   \\

&  &  &  &  &  &  &  & &\\

\hline
\end{array}$$

\end{table}
%
%
%
With solutions (\ref{sm3prRG-sol}) and (\ref{sols-aboveMI}) of RG equations  at corresponding energy scales, and taking into account the boundary conditions (\ref{bound-Yw})-(\ref{unif-MG}), we derive
\beq
\l \!\!\!\begin{array}{cccc}
  (b_1^{IG}-b_Y^{ZI}+b_{3'}^{\La'I}) , \!&\hs{-2mm} -b_1^{IG}  ,&\hs{-2mm} (b_{3'}^{II'}-b_{1'}^{II'}) ,  & \hs{-3mm}-2\pi \\
  (b_2^{IG}-b_w^{ZI}+b_{3'}^{\La'I}), \!&\hs{-2mm} -b_2^{IG} , &\hs{-2mm} (b_{3'}^{II'}-b_{2'}^{II'}) , &\hs{-3mm} -2\pi \\
  (b_3^{IG}-b_c^{ZI}) , &\hs{-2mm} -b_3^{IG} , \!&\hs{-2mm} 0 ,&\hs{-3mm} -2\pi  \\
   (b_{5'}^{I'G}-b_{3'}^{\La'I}) , \!&\hs{-2mm} -b_{5'}^{I'G} ,&\hs{-2mm}  (b_{5'}^{I'G}-b_{3'}^{II'}) , &\hs{-3mm} -2\pi
\end{array} \!\!\r \!\!
\l \!\!\!\begin{array}{c}
  \ln \fr{M_I}{M_Z} \\
  \ln \fr{M_G}{M_Z}  \\
   \ln \fr{{M_I}'}{M_I}  \\
  \al_G^{-1}
\end{array}\!\!\!\r \!\!=\!\!
\l \!\!\!\begin{array}{c}
  2\pi(\al_{3'}^{-1}(\La')\!-\!\al_Y^{-1})+b_{3'}^{\La'I}\ln \fr{\La'}{M_Z} \\
  2\pi(\al_{3'}^{-1}(\La')\!-\!\al_w^{-1})+b_{3'}^{\La'I}\ln \fr{\La'}{M_Z} \\
  -2\pi \al_c^{-1}\\
  -2\pi \al_{3'}^{-1}(\La')-b_{3'}^{\La'I}\ln \fr{\La'}{M_Z}
\end{array}\!\!\!\r ,
\la{calc-unif}
\eeq
where on the right-hand side of this equation the couplings $\al_{Y,w,c}$ are taken at scale $M_Z$.
The factors $b_i^{\mu_a\mu_b}$ (like $b_1^{IG}$, $b_{3'}^{\La'I}$ etc.) stand for effective $b$ factors corresponding to the energy interval
$\mu_a-\mu_b$ and can also include two-loop effects. All expressions and details are given in Appendix \ref{rg}.

\begin{figure}
\begin{center}
\leavevmode
\leavevmode
\vspace{-0.5cm}
\includegraphics{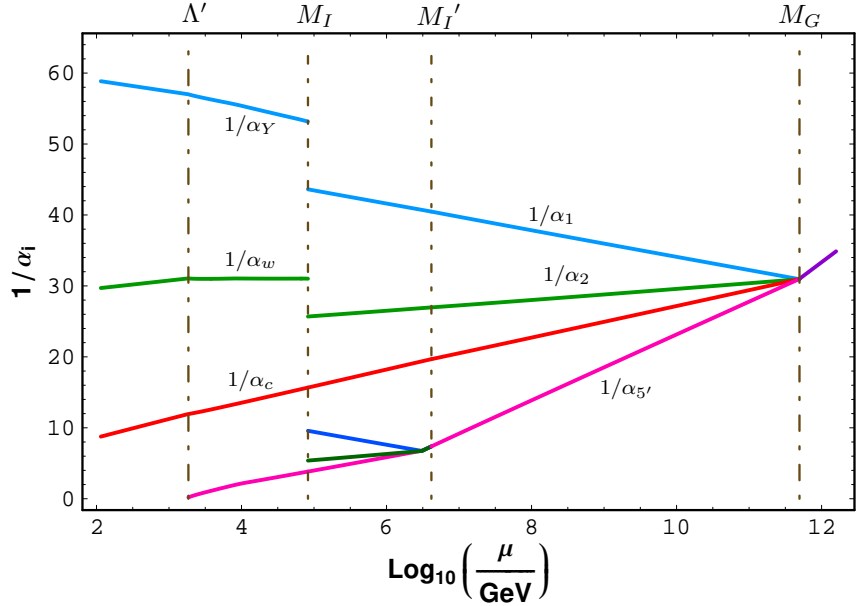}  
\end{center}
\vs{7.4cm}
\caption{Gauge coupling unification.
$\{\La', M_I, {M_I}', M_G \}\simeq  \{1800, 8.25\cdot \!10^4, 4.16\cdot \!10^6, 4.95\cdot \!10^{11}\}$~GeV and
$\al_G(M_G)\simeq 1/31$.}
\rput(11.8,4.5){\scriptsize $1/\al_{5'}$}
\rput(10.8,6.8){\scriptsize $1/\al_1$}
\rput(11,6){\scriptsize $1/\al_2$}
\rput(6.8,8){\scriptsize $1/\al_Y$}
\rput(6.8,6.2){\scriptsize $1/\al_w$}
\rput(6.8,4.6){\scriptsize $1/\al_c$}
\rput(6.05,9.5){\footnotesize $\La'$}
\rput(7.65,9.45){\footnotesize $M_I$}
\rput(9.3,9.45){\footnotesize ${M_I}'$}
\rput(14.1,9.45){\footnotesize $M_G$}
\label{fig3}
\end{figure}

From Eq. (\ref{calc-unif}) we can calculate $\{M_I,~ M_G,~{M_I}' ,~\al_G\}$ in terms of the remaining inputs.
For instance, a phenomenologically viable scenario is obtained when $SU(3)'$ confines at scale $\La'\sim 1$~TeV. Thus, we will take
$\La'\sim 1$~TeV and $\al_{3'}^{-1}(\La')\ll 1$.  In Table \ref{t:tab1} we give
selected input mass scales, leading to successful unification with
\beq
\{M_I,~{M_I}' ,~ M_G \}\simeq  \{8.25\cdot \!10^4, 4.16\cdot \!10^6, 4.95\cdot \!10^{11}\}~{\rm GeV}~,~~~
\al_G\simeq 1/31 ~.
\la{output}
\eeq
The corresponding picture of gauge coupling running is given in Fig. \ref{fig3}.
This result is obtained by solving RGs in the two-loop approximation. More details, including one- and two-loop RG factors at each relevant mass scale,
are given in Appendix \ref{rg}.

\section{Nucleon stability}
\la{Pdecay}

In this section we show that, although the GUT scale $M_G$ is relatively low (close to $5\cdot 10^{11}$~GeV), the nucleon's lifetime can be compatible
with current experimental bounds. In achieving this, a crucial role is played by lepton compositeness, because leptons have no direct
couplings with $X, Y$ gauge bosons of $SU(5)$. The baryon number violating $d=6$ operators, induced by integrating
out of the $X,Y$ bosons, are
\beq
\fr{g_X^2}{M_X^2}(\ov{u^c}_a\ga_{\mu}q_b^i)(\ov{d^c}_c\ga^{\mu}l^j)\ep^{abc}\ep_{ij}~,~~~~~~~
\fr{g_X^2}{M_X^2}(\ov{u^c}_a\ga_{\mu}q_b^i)(\ov{e^c}\ga^{\mu}q_c^j)\ep^{abc}\ep_{ij}~,
\la{d6-ops-B-L}
\eeq
where $g_X$ is the $SU(5)$ gauge coupling at scale $M_X$ (the mass of the $X,Y$ states). According to Eq. (\ref{int-lec}), the states $l, e^c$ contain light leptons $l_0, e^c_0$. Using this and going to the mass eigenstate basis [with
Eqs. (\ref{YUD-diag}) and (\ref{ckm})], from Eq. (\ref{d6-ops-B-L}) we get operators
$$
{\cal O}_{d6}^{(e^c)}=\fr{g_X^2}{M_X^2}{\cal C}^{(e^c)}_{\al\bt}(\ov{u^c}\ga_{\mu}u)(\ov{e^c_{\al}}\ga^{\mu}d_{\bt})~,~~~~~~~
{\cal O}_{d6}^{(e)}=\fr{g_X^2}{M_X^2}{\cal C}^{(e)}_{\al\bt}(\ov{u^c}\ga_{\mu}u)(\ov{d^c}_{\bt}\ga^{\mu}e_{\al})~,
$$
\beq
{\cal O}_{d6}^{(\nu)}=\fr{g_X^2}{M_X^2}{\cal C}^{(\nu)}_{\al\bt\ga}(\ov{u^c}\ga_{\mu}d_{\al})(\ov{d^c}_{\bt}\ga^{\mu}\nu_{\ga})~,
\la{d6-ops-B-L1}
\eeq
with
$$
{\cal C}^{(e^c)}_{\al\bt}=(R_u^\dag L_u^*)_{11}(R_e^\dag \tl{\mu}^*\fr{1}{M_{e^c\hat e^c}^*}L_u^*P_1^*V_{CKM})_{\al\bt}+
(R_u^\dag L_u^*P_1^*V_{CKM})_{1\bt}(R_e^\dag \tl{\mu}^*\fr{1}{M_{e^c\hat e^c}^*}L_u^*)_{\al 1} ~,
$$
$$
{\cal C}^{(e)}_{\al\bt}=(R_u^\dag L_u^*)_{11}(R_d^\dag\fr{1}{M_{\hat ll}}\hat{\mu}L_e^*)_{\bt\al} ~,
$$
\beq
{\cal C}^{(\nu)}_{\al\bt\ga}=(R_u^\dag L_u^*P_1^*V_{CKM})_{1\al}(R_d^\dag\fr{1}{M_{\hat ll}}\hat{\mu}L_e^*)_{\bt\ga}~,
\la{d6-C-coef}
\eeq
where in Eq. (\ref{d6-ops-B-L1}) we have suppressed the color indices.
Similar to quark Yukawa matrices, the charged lepton Yukawa matrix has been diagonalized by  transformation
$L_e^\dag Y_ER_e=Y_E^{\rm Diag}$. All fields in Eq. (\ref{d6-ops-B-L1}), are assumed to denote mass eigenstates. We have ignored the neutrino masses (having no relevance for the nucleon decay) and rotated the neutrino flavors $\nu_0=L_e^*\nu$ similar to
 the left-handed charged leptons  $e_0=L_e^*e$.

As we will show now, with proper selection of appropriate parameters (such as $\tl{\mu}\fr{1}{M_{e^c\hat e^c}}$,
$\fr{1}{M_{\hat ll}}\hat{\mu}$
and/or corresponding entries in some of unitary matrices), appearing in Eq. (\ref{d6-C-coef}), we can adequately
suppress nucleon decays within our model.\footnote{The importance of flavor dependence in $d=6$ nucleon decay was discussed in Refs.
\cite{d6-flavored} and \cite{d6-rotate}. As was shown \cite{d6-rotate},  in specific
circumstances, within GUTs one can suppress or even completely rotate away the  $d=6$ nucleon decays.}
Upon the selection of parameters, the constraint (\ref{constr-matr})
must be satisfied in order to obtain observed values of charged fermion masses. Introducing the notations
\beq
R_u^\dag L_u^*\equiv {\cal U}~,~~~R_d^\dag\fr{1}{M_{\hat ll}}\hat{\mu}L_e^*\equiv {\cal L}~,~~~
R_e^\dag \tl{\mu}^*\fr{1}{M_{e^c\hat e^c}^*}L_u^*\equiv {\cal R}~,
\la{d6-notat}
\eeq
the couplings in Eq. (\ref{d6-C-coef}) can be rewritten as
$$
{\cal C}^{(e^c)}_{\al\bt}={\cal U}_{11}({\cal R}P_1^*V_{CKM})_{\al\bt}+
({\cal U}P_1^*V_{CKM})_{1\bt}({\cal R})_{\al 1} ~,
$$
\beq
{\cal C}^{(e)}_{\al\bt}={\cal U}_{11}{\cal L}_{\bt\al} ~,~~~~
{\cal C}^{(\nu)}_{\al\bt\ga}=({\cal U}P_1^*V_{CKM})_{1\al}{\cal L}_{\bt\ga}~.
\la{d6-C-coef1}
\eeq
Since the matrices ${\cal U}, {\cal L}$ and ${\cal R}$ are not fixed yet, for their structures we will make the selection
\beq
{\cal U}_{11}=0,~~~
{\cal L}=\left(\!\!
  \begin{array}{ccc}
    \ep_1& \ep_2 & \ep_3 \\
    \tm & \tm & \tm \\
    \tm & \tm & \tm \\
  \end{array}\!\!
\right),~~~
{\cal R}=\left(\!\!
  \begin{array}{ccc}
    0& \tm & \tm \\
    0 & \tm & \tm \\
    \tm & \tm & \tm \\
  \end{array}\!\!
\right),
\la{form-ULR}
\eeq
where $\tm $ stands for some nonzero entry. With this structure we see that   for $\al ,\bt =1,2$ we have
 ${\cal C}^{(e^c)}_{\al \bt}={\cal C}^{(e)}_{\al \bt }=0$, and therefore
nucleon decays with emission of the charged leptons do not take place.  With one more selection, we will be able to eliminate some nucleon decay
modes (but not all) with neutrino emissions.  We can impose one more condition, involving
${\cal U}_{12}$ and ${\cal U}_{13}$ entries of ${\cal U}$, in such a way as
to have $({\cal U}P_1^*V_{CKM})_{11}=0$. The latter, in expanded form, reads
\beq
({\cal U}P_1^*V_{CKM})_{11}={\cal U}_{12}e^{-i\om_2}V_{cd}+{\cal U}_{13}e^{-i\om_3}V_{td}=0~,~~~ \Longrightarrow
{\cal U}_{12}e^{-i\om_2}=-\fr{V_{td}}{V_{cd}}{\cal U}_{13}e^{-i\om_3}
\la{cond-U}
\eeq
and leads to ${\cal C}^{(\nu)}_{12\ga}={\cal C}^{(\nu)}_{11\ga}=0$. Thus, the decays $p\to \bar{\nu}\pi^+, n\to \bar{\nu}\pi^0, n\to \bar{\nu}\eta $
do not take place. Nonvanishing relevant ${\cal C}^{(\nu)}$ couplings are  ${\cal C}^{(\nu)}_{21\ga }$, which, taking into account
Eqs. (\ref{form-ULR}) and (\ref{cond-U}), are
\beq
{\cal C}^{(\nu)}_{21\ga}=({\cal U}P_1^*V_{CKM})_{12}\ep_{\ga}=\ep_{\ga}{\cal U}_{13}e^{-i\om_3}\fr{V_{ts}V_{cd}-V_{td}V_{cs}}{V_{cd}}\simeq
\ep_{\ga}{\cal U}_{13}e^{-i\om_3}\fr{s_{13}e^{i\de}}{V_{cd}}~,
\la{Cnu-21}
\eeq
 where in last step we have used standard parametrization of the CKM matrix.
Since the matrix ${\cal U}$ is unitary, due to selection ${\cal U}_{11}=0$ and the unitarity condition,
we will have $|{\cal U}_{12}|^2+|{\cal U}_{13}|^2=1$. With this, by Eq. (\ref{cond-U}) and using central values \cite{Beringer:1900zz}
of CKM matrix elements,
we obtain $|{\cal U}_{12}|\simeq 0.038, |{\cal U}_{13}|\simeq 1$ and $|\fr{s_{13}}{V_{cd}}|=|\fr{V_{ub}}{V_{cd}}|\simeq 1.56\cdot 10^{-2}$.
These give
$|{\cal C}^{(\nu)}_{21\ga}|\simeq 1.56\cdot 10^{-2}|\ep_{\ga}|$.
Taking into account all this, for expressions of $p\to \bar{\nu}K^+$ and $n\to \bar{\nu}K^0$  decay widths, we obtain \cite{p-decay-widths}
\beq
\Ga (p\to \bar{\nu}K^+)\!\simeq \!\Ga (n\to \bar{\nu}K^0)\!=\!\fr{(m_p^2-m_K^2)^2}{32\pi f_{\pi}^2m_p^3}\!
\l \!1\!+\!\fr{m_p}{3m_B}(D+3F)\!\!\r^2\!\!\l \!\fr{g_X}{M_X^2}A_R|\al_H|\!\!\r^2 \!\!\!\cdot 2.43\cdot 10^{-4}
\!\sum_{\ga=1}^3|\ep_{\ga}|^2
\la{pnuK-width}
\eeq
where $|\al_H|=0.012~{\rm GeV}^3$ is a hadronic matrix element and $A_R=A_LA_S^l\simeq 1.48$ takes into account long- ($A_L\simeq 1.25$) and
short-distance ($A_S^l\simeq 1.18$)  renormalization effects (see Refs. \cite{Nihei:1994tx} and \cite{Buras:1977yy}, respectively. Some details
of the calculation of $A_S^l$, within our model, are given in Appendix \ref{d6RG}).
To satisfy current experimental bound $\tau_p^{exp}(p\to \bar{\nu}K^+)\stackrel{<}{_\sim}5.9\cdot 10^{33}$ years \cite{Babu:2013jba}, for
$M_X\simeq 5\cdot 10^{11}$~GeV and $\al_X\simeq 1/31$, we need to have $\sqrt{|\ep_1|^2+|\ep_3|^2+|\ep_3|^2}\stackrel{<}{_\sim} 4.8\cdot 10^{-6}$. This selection of parameters is fully consistent with the charged fermion masses. Note, that with Eq. (\ref{form-ULR}) there is no conflict with the constraint of Eq. (\ref{constr-matr}).
We can lower values of $|\ep_{\ga}|$; however, there is a low bound dictated from this constraint. With
$|{\rm Det} (\tl{\mu}\fr{1}{M_{e^c\hat e^c}})|\cdot |{\rm Det}(\fr{1}{M_{\hat ll}}\hat{\mu})|=|{\rm Det}({\cal L})|\cdot |{\rm Det}({\cal R})|\sim 10^{-8}$, the
lowest value can be $|\ep_{\ga}|\sim 10^{-8}$, obtained with $|{\rm Det}({\cal R})|\sim 1$. More natural would be to have
 $|{\rm Det}({\cal R})|\stackrel{<}{_\sim} 10^{-2}$,
which suggests  $|{\rm Det}({\cal L})|\stackrel{>}{_\sim}  10^{-6}$, and therefore
$\sqrt{|\ep_1|^2+|\ep_3|^2+|\ep_3|^2}\stackrel{>}{_\sim } \sqrt{3}\cdot 10^{-6}$. This dictates an upper bound for the proton lifetime
$\tau_p=\tau(p\to \bar{\nu}K^+)\stackrel{<}{_\sim } 5\cdot 10^{34}$ years and will allow us to test the model in the future \cite{Babu:2013jba}.

Besides $X, Y$ gauge boson mediated operators, there are $d=6$ operators generated by the exchange of colored triplet scalar $T_{H}$.
 From the couplings of Eq. (\ref{from-LY}), we can see that the integration of $T_H$ induces baryon number violating
 $\fr{1}{M_{T_H}^2}(q^TC_{qq}q)(q^TC_{ql}l)$ and $\fr{1}{M_{T_H}^2}(u^cC_{u^ce^c}e^c)(u^cC_{u^cd^c}d^c)$ operators, which lead to the couplings
 $\fr{1}{M_{T_H}^2}(q^TC_{qq}q)(q^TC_{ql}\fr{1}{M_{\hat ll}}\hat{\mu}l_0)$ and
 $\fr{1}{M_{T_H}^2}(u^cC_{u^ce^c}\fr{1}{M_{e^c\hat e^c}^T}\tl{\mu}^Te^c_0)(u^cC_{u^cd^c}d^c)$.
 Couplings $C_{ab}$ appearing in these operators are independent from Yukawa matrices, and proper suppression of relevant terms is possible
 [similar to the case of couplings in Eq. (\ref{d6-C-coef1})], leaving fermion masses and a mixing pattern consistent with experiments. To make a
 more definite statement about the nucleon lifetime, one has to study in detail the structure of Yukawa matrices.
  In  this respect, extension with flavor symmetries
 is a motivated framework and can play a crucial role in generating the desirable Yukawa textures [guaranteeing the forms given in  Eq. (\ref{form-ULR})].
Preserving these issues for being addressed elsewhere, let us move to the next section.

\section{Various phenomenological constraints and implications}
\la{impl}

In this section we discuss and summarize some peculiarities,  phenomenological implications of our model, and constraints
needed to be satisfied in order to be consistent with experiments. Also, we list issues opening prospects for further investigations
within presented scenario:

\vs{0.2cm}
{\bf (i)} The discovery of the Higgs boson  \cite{Aad:2012tfa}, with mass $\approx 126$~GeV, revealed that the Standard Model suffers from vacuum instability.
Detailed analysis has shown \cite{Buttazzo:2013uya} that, due to RG, the Higgs
self-coupling  becomes negative near the scale $\sim 10^{10}$~GeV.
If the Higgs field is insured to remain in the EW vacuum, the problem perhaps is not as severe. However, with an inflationary universe
with the Hubble parameter$\gg 10^{10}$~GeV (preferred by the recent BICEP2 measurement \cite{Ade:2014xna}), the EW vacuum can be easily
destabilized by the Higgs's move/tunneling to the "true" anti-de Sitter (AdS) vacuum  \cite{Hook:2014uia}.
Whether AdS domains take over or crunch depends on the details of inflation, the reheating
process, nonminimal Higgs/inflaton couplings, etc. (a detailed overview of these questions can be found in Refs. \cite{Espinosa:2007qp} and \cite{Hook:2014uia}).
 While these and related issues need more investigation,
to be on t safe sehide, it is desirable to have a model with positive $\lam_h$ at all energy scales (up to the $M_{\rm Pl}$).

Since within our model above the $\La'$ scale new states  appear, this problem can be avoided. As was mentioned in Sec. \ref{55model},
in our model a light SM doublet $h$ dominantly comes from the $H$-plet. The coupling $\lam_H(H^\dag H)^2$ gives the self-interaction term $\lam_h(h^\dag h)^2$
(with $\lam_h\approx \lam_H$ at the GUT scale). The running of  $\lam_h$
will be given by
$$
16\pi^2\fr{d}{dt}\lam_h=\bt_{\lam_h}^{SM}+\De \bt_{\lam_h}~,
$$
where $\bt_{\lam_h}^{SM}$ corresponds to the SM part, while $\De \bt_{\lam_h}$ accounts for new contributions.
 Since the $H$-plet in the potential (\ref{V-terms}) has additional interaction terms, some of those couplings can help to increase $\lam_h$.
 For instance, the couplings $\lam_{1H\Phi }, \lam_{2H\Phi }$, $\hat h$, etc., contribute as
$$
\De \bt_{\lam_h}\approx \fr{(\lam_{1H\Phi })^2}{25}\left [ 9\te (\mu \!-\!M_{TT'})+6\te (\mu \!-\!M_{DT'})+6\te (\mu \!-\!M_{TD'})+4
\te (\mu \!-\!M_{DD'})\right ]
$$
\beq
\fr{(\lam_{2H\Phi })^2}{10}\left [3\te (\mu \!-\!M_{DT'})+2\te (\mu \!-\!M_{DD'})  \right ]+3\hat h^2\te (\mu \!-\!M_{T_{H'}})+\cdots
\la{impr-lamh}
\eeq
Detailed analysis requires numerical studies by solving the system of coupled  RG equations (involving multiple couplings\footnote{For methods
studying the stability of multifield potentials, see Refs. \cite{Sher:1988mj} and \cite{Chakrabortty:2013mha} and references therein.}).
While this is beyond the scope of this work, we see that
due to positive contributions (see above) into the $\bt$ function, there is potential to prevent $\lam_h$ becoming negative all the way
up to the Planck scale.

\vs{0.2cm}
{\bf (ii)} Since in our model leptons are composite, there will be additional contributions to
their anomalous magnetic moment, given by \cite{Barbieri:1980aq}
\beq
\de a_{\al}\sim \l \fr{m_{e_{\al}}}{\La'}\r^2 ~.
\la{e-AMM}
\eeq
Current experimental measurements \cite{Beringer:1900zz} of the muon anomalous magnetic moment give $\De a_{\mu }^{\rm exp}\approx 6\cdot 10^{-10}$.
This, having in mind a possible range $\sim (1/5-1)$ of an undetermined prefactor in the expression of Eq. (\ref{e-AMM}),
constrains the scale $\La'$ from below: $\La'\stackrel{>}{_\sim } (1.8-4.3)$~TeV. The selected value of $\La'$, within our model ($\La'=1851$~GeV),
 fits well with this bound.\footnote{In fact, this new contribution to $a_{\mu}$ has the potential of  resolving a $3-4\si$ discrepancy \cite{Beringer:1900zz}
(if it will persist in the future) between the theory and experiment \cite{Passera:2010ev}.}
The value of $\de a_e$ is more suppressed (for  $\La'\simeq 1.8$~TeV,  we get $\de a_e\sim 10^{-13}$) and is compatible with experiments
($\De a_e^{\rm exp}\approx 2.7\cdot 10^{-13}$). Planned measurements \cite{LeeRoberts:2011zz}
with reduced  uncertainties  will provide severe constraints and  test the viability of the proposed scenario.

Similarly, having flavor violating couplings at the level of constituents (i.e., in the sector of $SU(3)'$ fermions $\hat q, \hat u^c, \hat d^c$),
the new contribution in $e_{\al}\to e_{\bt}\ga $ rare decay processes will emerge.  For instance, the contribution in the $\mu \to e\ga $ transition amplitude will be
 $\sim \lam_{12}\fr{m_{\mu}}{(\La')^2}$, where $\lam_{12}$ is (unknown) flavor violating coupling coming from the Yukawa sector of $\hat q, \hat u^c, \hat d^c$.
 This gives $Br(\mu\to e\ga)\sim \lam_{12}^2(\fr{M_W}{\La'})^4$, and for $\La'\simeq 1.8$~TeV  the constraint
 $\lam_{12}\stackrel{<}{_\sim }4\cdot 10^{-4}$ should be satisfied in order to be consistent with the latest experimental
 limit $Br^{\rm exp}(\mu\to e\ga)<5.7\cdot 10^{-13}$ \cite{Adam:2013mnn}.

\vs{0.2cm}
{\bf (iii)}
As was mentioned in Sec. \ref{fermion-mass} (and will be discussed  also in Appendix \ref{comp-anom}), the matter sector of $SU(3)'$ symmetry (ignoring EW
and Yukawa interactions) possesses $G_f^{(6)}$ chiral symmetry with sextets $6_L\sim \hat q_{\al }$ and $6_R\sim \hat q^c_{\al }$
[see Eqs. (\ref{Gf6}) and (\ref{Gf-transf})]. The breaking of this chiral symmetry proceeds by several steps. At the first stage, at
scale $\La'\approx 1.8$~TeV, the condensates $\lan 6_L6_LT_{H'}^\dag \ran \sim \lan 6_R6_RT_{H'} \ran \sim \La'$ break the $G_f^{(6)}$. However,
these condensates preserve SM gauge symmetry. At the next stage (of chiral symmetry breaking), the condensate $\lan 6_L6_R\ran \equiv F_{\pi'}$,
together with the Higgs VEV $\lan h\ran \equiv v_h$, contributes to the EW symmetry breaking. The $F_{\pi'}$ denotes the decay constant of the (techni) $\pi'$
meson and should satisfy $v_h^2+F_{\pi'}^2=(246.2~{\rm GeV})^2$. With the light (very SM-like) Higgs boson  mainly residing in  $h$ and with
$F_{\pi'}\stackrel{<}{_\sim }0.2v_h$, the $h$'s signal will be very compatible with LHC data \cite{Carone:2012cd}.
Since the low-energy potential would involve VEVs  $\lan 6_L6_LT_{H'}^\dag \ran , \lan 6_R6_RT_{H'} \ran , F_{\pi'}$ and $v_h$, obtaining  mild hierarchy
$\fr{F_{\pi'}}{\La'}\stackrel{<}{_\sim }1/40$ will be possible by proper selection (not by severe fine-tunings) of parameters from perturbative  and nonperturbative (effective) potentials.
The situation here (i.e., the symmetry breaking pattern, potential (being quite involved because of these VEVs), etc.) will differ from case obtained
within  QCD with $SU(n)_L\tm SU(n)_R$ chiral symmetry and with the $\lan n_L\tm n_R\ran $ condensate only  \cite{Coleman:1980mx}.
Moreover, the hierarchy between the confinement scale and the decay constant can have some dynamical origin
(see, e.g., Refs.\footnote{If a conformal window is realized, the value of $F_{\pi'}$ can be more reduced \cite{DelDebbio:2010zz}.}
 \cite{Appelquist:2009ka}).
Without addressing these details, our approach is rather phenomenological, with the assumption $F_{\pi'}/v_h\stackrel{<}{_\sim }0.2$
and  $h$ being the Higgs boson (with mass $\approx 126$~GeV), such that there is allowed a window for a heavier $\pi'$ state
and the model is compatible with current experiments \cite{Belanger:2013xza}.
Models with partially composite Higgs, in which the light Higgs doublet has some ed-mixture of a composite (technipion $\pi'$) state,
with various interesting implications (including necessary constraints, limits, and compatibility with LHC data), were studied  in Ref.
\cite{Carone:2012cd}.
As mentioned in Sec.  \ref{unif}, it is possible to have unification with the symmetry breaking pattern and the spectrum of intermediate states
that give larger values of $\La'$ (even with  $\La'\sim 10^5$~GeV).
However, in such a case, the value of $F_{\pi'}$ would be also large,
and it would be impossible to bring $F_{\pi'}$ to the low value even with
fine-tuning.
This would mean that the EW symmetry breaking scale would be also large. That is why such a possibility has not been considered.

In addition, it is rather generic that the model with composite leptons will be accompanied with excited massive leptons (lepton resonances). Current
experiments have placed low bounds on masses of the excited electron and muon to be heavier than $\sim 1.8$~TeV.
This scale is close to the value of $\La'$ we have chosen within our model, and will allow us to test the lepton substructure
\cite{Eichten:1983hw} hopefully in the not-far future.
Details, related to these issues, deserve separate investigations.

\vs{0.2cm}
{\bf (iv)}
Since the condensate $\lan 6_L6_R\ran =F_{\pi'}$, by some amount, can contribute to the chiral [of the $SU(3)'$ strong sector] and
EW symmetry breaking, the scenario shares some properties of hybrid technicolor models with fundamental Higgs states. Moreover, together with
technipion $\pi'$, near the $\La'$ scale, there will be technimeson states
$\rho_T, \om_T$, etc., with peculiar signatures \cite{Hod:2013cba}, \cite{Eichten:2012hs},
which can be probed by collider experiments.

\vs{0.2cm}
{\bf (v)}
Because the new states around and above the $\La'\approx 1.8$~TeV scale, there will be additional corrections to the EW precision
parameters $T, S, U$ etc. While because strong dynamics near the $\La'$ scale, the accurate calculations require some effort,
the symmetry arguments provide a good estimate of the additional corrections $\De T, \De S$, etc. One can easily notice that the isospin
breaking effects are suppressed in the sector of additional states. Therefore the mass splittings between doublet components
of the additional states will be suppressed (i.e. $\De M\ll M$) and  pieces $\De T_f, \De T_s$ of $\De T=\De T_f+\De T_s$ will be
given as \cite{Einhorn:1981cy}
\beq
\De T_f\simeq \fr{N_f}{12\pi s_W^2}\l \fr{\De M_f}{m_W}\r^2~,~~~~~\De T_s\simeq \fr{N_s}{24\pi s_W^2}\l \fr{\De M_s}{m_W}\r^2~,
\la{T-fs}
\eeq
where subscripts $f$ and $s$ stand for fermions, and scalars, respectively and $N_f, N_s$ account for the multiplicity [or dimension with respect to the group different from
$SU(2)_w$] of the corresponding doublet state.
One can easily verify that within our model in the sector of extra vectorlike $(\hat{l}+l)_{\al}$ states the mass splitting between doublet
components is suppressed as $\De M_{\hat ll}^{(\al )}\stackrel{<}{_\sim }\fr{v_h^2}{M_{\hat ll}^{(\al )}}$. This, according to Eq. (\ref{T-fs})
and Table \ref{t:tab1}, gives the negligible contribution:
$\De T_{\hat ll}\stackrel{<}{_\sim }\fr{2 \cdot 2}{12\pi s_W^2}v_h^4/(m_WM_{\hat ll}^{(1)})^2\sim 10^{-5}$.
Within the fragments  of the scalar $\Phi $, the lightest is $\Phi_{DT'}$ with mass $M_{DT'}\simeq 8.3$~TeV. Splitting between the doublet components
comes from the potential term $\fr{\lam_{2H\Phi }}{\sqrt{10}} \!H^\dag \Phi  \Phi^\dag \!H $, giving
$\De M_{DT'}\simeq \lam_{2H\Phi }v_h^2/(4\sqrt{10}M_{DT'})$. This, according to Eq. (\ref{T-fs}), causes enough suppression:
$\De T_{DT'}\simeq \fr{3}{24\pi s_W^2}\lam_{4H\Phi }^2v_h^4/(2\sqrt{10}M_{DT'}m_W)^2\stackrel{<}{_\sim }2\cdot 10^{-5}$
(for $\lam_{2H\Phi }\stackrel{<}{_\sim }1.5$). As pointed out above, besides the fundamental Higgs doublet ($h$), which dominantly
includes SM Higgs, there is a composite doublet ($\pi'$ - similar to technicolor models) with suppressed VEV - $F_{\pi'}$.
Contribution of this extra doublet, into the $T$ parameter, is estimated to be
\beq
\De T_{\pi'}\approx \fr{1}{24\pi s_W^2}\!\l \!\fr{\De M_{\pi'}}{m_W}\!\r^2 -\fr{c_W^2}{4\pi}c_{\pi'}^2\ln \fr{M_{\pi'}^2}{m_Z^2} ~,
\la{T-pi1}
\eeq
where the first term is due to the mass splitting $\De M_{\pi'}(\sim v_h^2/(4M_{\pi'})$) between doublet components of $\pi'$, while second term emerges due to the VEV $\lan \pi'\ran =F_{\pi'}$
with $c_{\pi'}\approx 2m_Z^2F_{\pi'}/(M_{\pi'}^2v_h)$ (where $F_{\pi'}\stackrel{<}{_\sim}0.2v_h$). This contribution is also small
($\De T_{\pi'}\approx 2\cdot 10^{-3}$) for $M_{\pi'}\sim 1$~TeV. Since $\pi'$ is a composite state, due to the strong dynamics, special care is needed to derive a
more accurate result (as was done in Ref. \cite{Peskin:1991sw}
for models with a single composite Higgs performing proper matching at different energy scales).
 However, since $\De T_{\pi'}$ is protected by isospin symmetry,
we limit ourselves to the estimates performed here. Moreover,  the source of the isospin breaking in the strong $SU(3)'$ sector is
$F_{\pi'}\stackrel{<}{_\sim}0.2v_h$, causing the mass splitting between composite "technihadrons" (denoted collectively as $\{ \rho' \}$)
of $\De M_{\rho'}\sim F_{\pi'}^2/M_{\rho'}$. This, for $M_{\rho'}\sim \La'$, would give the
 correction $\De T_{\rho'}\sim \fr{1}{12\pi s_W^2}F_{\pi'}^4/(m_WM_{\rho'})^2\stackrel{<}{_\sim} 10^{-5}$. Note that the direct isospin (custodial symmetry) breaking within $\hat{q}_{\al}$ states is much more suppressed (we have no direct EW symmetry breaking in the Yukawa sector of $\hat{q}, \hat{u}^c$,
 and $\hat{d}^c$ states) and thus conclude that within the considered scenario extra corrections to the $T$ parameter are under control.

Let us now give the estimate of the additional contributions into the $S$ parameter. Contributions to this parameter from the additional vectorlike
  $(\hat{l}+l)_{\al}$,  $(\hat{e}^c+e^c)_{\al}$ states decouple  \cite{Lavoura:1992np} and are estimated to be
  $\De S_{\hat ll}\sim \De S_{\hat{e}^c e^c}\stackrel{<}{_\sim }\fr{1}{4\pi}\fr{v_h^2}{(M_{\hat ll}^{(1)})^2}\ln \fr{M_{\hat ll}^{(1)}}{m_{\tau}}\sim 10^{-5}$.
  The contribution from the scalar $\Phi_{DT'}$ is
  $\De S_{DT'}\simeq \fr{3}{6\pi}\De M_{DT'}/M_{DT'}\simeq \lam_{2H\Phi }v_h^2/(8\pi \sqrt{10}M_{DT'}^2)\stackrel{<}{_\sim } 2\cdot 10^{-5}$,
  also suppressed, as expected.
The contribution of extra (heavy $\pi'$) composite doublet is
\beq
\De S_{\pi'}\approx \fr{1}{6\pi }\fr{\De M_{\pi'}}{M_{\pi'}}+\fr{1}{6\pi }c_{\pi'}^2\ln \fr{M_{\pi'}}{m_h}~,
\la{S-pi1}
\eeq
where first term is due to the splitting of the doublet components, while second term comes from the VEV $\lan \pi'\ran =F_{\pi'}$.
With $\De M_{\pi'}\sim v_h^2/(4M_{\pi'})$ and $M_{\pi'}\stackrel{>}{_\sim }1$~TeV Eq. (\ref{S-pi1}) gives $\De S_{\pi'}\stackrel{<}{_\sim } 10^{-3}$.
Similarly suppressed contributions would arise from the techni$-{\rho'}$ hadrons:
$\De S_{\rho'}\sim \fr{1}{6\pi}\De M_{\rho'}/M_{\rho'}\sim \fr{1}{6\pi}F_{\pi'}^2/M_{\rho'}^2\stackrel{<}{_\sim }4\cdot 10^{-5}$ (for $M_{\rho'}\sim \La'$).

As far as the contribution from the matter states $\hat{q}, \hat{u}^c,  \hat{d}^c$ are concerned, since their masses are too suppressed,
in the chiral limit $\fr{m_f}{m_Z}\to 0$, we can use the expression \cite{Einhorn:1981cy}
$$
\De S_f\to \fr{N_fY_f}{6\pi}\l -2\ln \fr{x_1}{x_2}+G(x_1)-G(x_2)\r ~,
$$
\beq
{\rm with}~~~~~~
G(x)=-4{\rm arc\!\tanh}\fr{1}{\sqrt{1-4x}}~,~~~~~x_i=\fr{m_{fi}^2}{m_Z^2}~,
\la{S-f}
\eeq
where $m_{f1,2}$ are masses of the components ofthe  $f$ fermion with hypercharge $Y_f$. Verifying that in the limit $x\to 0$ the function $G(x)$ goes to
$2\ln x$, we see that expression for $\De S_f$ in Eq. (\ref{S-f}) vanishes. Moreover, new contributions to the $U$ parameter
are more suppressed. For instance, the contribution due to the $\pi'$ is
 \beq
 \De U_{\pi'}\approx \fr{1}{15\pi}\l \!\fr{\De M_{\pi'}}{M_{\pi'}}\!\r^2 \!\!-\fr{1}{12\pi}c_{\pi}^2\fr{\De M_{\pi'}}{M_{\pi'}}~,
 \la{U-pi1}
 \eeq
 which for $M_{\pi'}\sim 1$~TeV, $F_{\pi'}\stackrel{<}{_\sim }0.2v_h$ becomes  $\De U_{\pi'}\stackrel{<}{_\sim }5\cdot 10^{-6}$.
All other new contributions to the $U$ are also more suppressed than the corresponding $\De S$ and $\De T$.
 This is understandable since $U$  is related to the effective operator with a dimension higher  than those of $S$ and $T$.
All these allow us to conclude that new contributions to the EW precision parameters are well below the current experimental bounds \cite{Baak:2011ze}.

\vs{0.2cm}
{\bf (vi)}
Within the proposed model,  spontaneous breaking of two non-Abelian groups $SU(5)\tm SU(5)'$ and discrete $D_2$ parity
will give monopole and domain wall solutions, respectively.  Since the symmetry breaking scales are relatively
low ($\stackrel{<}{_\sim }5\cdot 10^{11}$~GeV),  the inflation would not dilute number densities of these topological defects in a straightforward way.
Thus, one can think of alternative solutions. For instance, as it was shown in Refs. \cite{Dvali:1995cc}, within models with a certain field content and
couplings, it is possible that symmetry restoration cannot happen for arbitrary high temperatures. This would avoid the phase transitions
(which usually cause the formation of topological defects). Moreover, by proper selection of the model parameters, it is
possible to suppress the thermal production rates of the topological defects (for detailed discussions, see the last two works of Ref. \cite{Dvali:1995cc}).
From this viewpoint, our model with a multiscalar sector and various couplings has potential to avoid  domain wall and monopole problems.
Thus, it is inviting to investigate the parameter space and see how desirable ranges are compatible with those needed values appearing
in Eq. (\ref{impr-lamh}) (for "improving" the running of $\lam_h$).

To cure problems related with topological defects, also other different noninflationary solutions have been proposed
\cite{Langacker:1980kd}, and one (if not all) of them could be invoked as well.

Certainly, these and other cosmological implications, of the presented scenario, deserve separate investigations.

 \vs{0.2cm}
At the end let us note  that it would be interesting to build a supersymmetric extension of the considered $SU(5)\tm SU(5)'\tm D_2$ GUT and study related phenomenology. These and related issues will be addressed elsewhere.

\subsubsection*{Acknowledgments}

I am grateful to K.S. Babu, J. Chkareuli, I. Gogoladze, and S. Raby  for useful comments and discussions.
The partial support from Shota Rustaveli National Science Foundation (Contracts No. 31/89 and No. 03/113) are kindly acknowledged.
 I would like to thank CETUP* (Center for Theoretical Underground Physics and Related Areas), supported by the US Department of Energy under Grant No. DE-SC0010137 and by the US National Science Foundation under Grant No. PHY-1342611, for its hospitality and partial support during the 2013 Summer Program.
 I also thank Barbara Szczerbinska for providing a stimulating atmosphere in Deadwood during this program.

\appendix

\renewcommand{\theequation}{A.\arabic{equation}}\setcounter{equation}{0}

\section{Composite leptons and anomaly matching}
\la{comp-anom}

Here we demonstrate how the composite leptons emerge within our scenario and also discuss anomaly matching conditions.
As was noted in Sec. \ref{fermion-mass}, the sector of $\hat q, \hat u^c$, and $\hat d^c$ states have $G_f^{(6)}$ chiral symmetry [see Eq. (\ref{Gf6})]
with the transformation properties of these states given in Eq. (\ref{Gf-transf}). At scale $SU(3)'$ interaction becomes strong, and the $G_f^{(6)}$ symmetry breaking condensates can be formed. The chiral symmetry breaking can proceed through several steps, and at each level the formed composite states should satisfy
anomaly matching conditions  \cite{hooft}.

 The bilinear [$SU(3)'$-invariant]
condensate can be $\lan 6_L\tm 6_R\ran =F_{\pi'}$, with corresponding breaking scale $F_{\pi'}$. As was shown in Ref. \cite{Coleman:1980mx}, with only
fundamental states, the chiral symmetry $SU(n)_L\tm SU(n)_R$ will be broken down to the diagonal $SU(n)_{L+R}$ symmetry. Since in our case $F_{\pi'}$
also contributes to EW symmetry breaking, we have a bound $F_{\pi'}\stackrel{<}{\sim }100$~GeV. This scale, in comparison with $\La'\sim $ few$\tm $TeV,
can be ignored at the first stage. Moreover, in our case, light $SU(3)'$ nonsinglet field content is reacher, and the chiral symmetry breaking pattern
is also different. Other $SU(3)'$ invariant condensates, including matter bilinears, are
\beq
\lan 6_L6_LT_{H'}^\dag \ran ~~~~~~~{\rm and} ~~~~~~~\lan 6_R6_RT_{H'} \ran ~.
\la{condens-qqT}
\eeq
Note, that the product of $SU(6)$ sextets gives either symmetric or antisymmetric representations ($6\tm 6=15_A+21_S$), but due to $SU(3)'$
contractions, in Eq. (\ref{condens-qqT}) the antisymmetric $15$-plets (i.e. $15_L$ and $15_R$) participate. The condensates (\ref{condens-qqT})
transform as $15_L$ and $15_R$ under $SU(6)_L$ and $SU(6)_R$, respectively, and therefore break these symmetries.
A possible breaking channel is
\beq
SU(6)_L\to SU(4)_L\tm SU(2)'_L\equiv G_L^{(4,2)}, ~~~~~~~SU(6)_R\to SU(4)_R\tm SU(2)'_R\equiv G_R^{(4,2)} .
\la{SU6s-br}
\eeq
Indeed, with respect to $G_L^{(4,2)}$ and $G_R^{(4,2)}$, the $15_L$ and $15_R$ decompose as
$$
SU(6)_L\to G_L^{(4,2)} :~~~15_L=(1,1)_L+(6,1)_L+(4,2)_L~,
$$
\beq
SU(6)_R\to G_R^{(4,2)} :~~~15_R=(1,1)_R+(6,1)_R+(4,2)_R~,
\la{15-dec}
\eeq
and the VEVs $\lan (1,1)_L\ran $ and $\lan (1,1)_R\ran $ leave $G_L^{(4,2)}\tm G_R^{(4,2)}$ chiral symmetry unbroken.
The singlet components ($\lan (1,1)_L\ran $ and $\lan (1,1)_R\ran $) from Eq. (\ref{condens-qqT}) are
$\fr{1}{2}\lan \hat q\hat qT_{H'}^\dag \ran =\lan \hat u\hat dT_{H'}^\dag \ran$ and $\lan \hat u^c\hat d^c T_{H'}\ran$ combinations, which  {\it leave}
$G_{SM}$ gauge symmetry unbroken. Therefore, the values of these condensates can be $\sim $few$\cdot $TeV($\sim \La'$) without causing any
phenomenological difficulties. Thus, as the first stage of the chiral symmetry breaking, we stick to the channel
\beq
G_f^{(6)}\stackrel{\La'}{_{\lrarr}} G_L^{(4,2)}\tm G_R^{(4,2)}\tm U(1)_{B'} ~,
\la{66-4242-br}
\eeq
with
\beq
\lan 6_L6_LT_{H'}^\dag \ran =\lan \hat u\hat dT_{H'}^\dag \ran \sim \La' ,~~~~
\lan 6_R6_RT_{H'} \ran =\lan \hat u^c\hat d^c T_{H'}\ran \sim \La' ~.
\la{condens-SM}
\eeq
The $SU(6)_{L,R}$ sextets under $G^{(4,2)}_{L,R}$ are decomposed as $6_L=(4,1)_L+(1,2)_L$ and $6_R=(4,1)_R+(1,2)_R$, respectively.
If composite objects are picked up as $(4',1)_{L,R}\subset [(4,1)_{L,R}]^3$ and $(1,2')_{L,R}\subset [(1,2)_{L,R}]^3$, then one can easily check out
that the anomalies (of initial and composite states) indeed match and $(4',1)_{L,R}$ and $(1,2')_{L,R}$ can be identified with three families of leptons
plus three states of right-handed/sterile neutrinos. For demonstrating all these, it is more convenient to work in a different basis. That would also make it simpler
to identify composite states.

As it is well known (and in our case turns out more useful), one can  describe the $SU(6)$  symmetry (and its representations as well) by
its  special subgroup ("S-subgroup" \cite{Slansky:1981yr}) $SU(3)_f\otimes SU(2)\subset SU(6)$. In our case,
\beq
SU(6)_L\supset SU(3)_{fL}\otimes SU(2)_L ~,~~~~SU(6)_R\supset SU(3)_{fR}\otimes SU(2)_R ~.
\la{max-S-sub}
\eeq
Under these S subgroups, the sextets decompose as \footnote{Similar to the description of three-flavor QCD with $(u, d, s)$ spin-1/2 states,
either by the sextet of $SU(6)$ or by $(3,2)$ of  $SU(3)_f\tm SU(2)_s$ - the Wigner-Weyl realization of the $SU(6)$ chiral symmetry.
Here, however, $SU(2)_s$ stands for the spin group
and $SU(3)_f$ for the flavor. In our case of Eq. (\ref{max-S-sub}), $SU(2)$ factors act like  isospin rotations
relating $\hat u_{\al}$ and $\hat d_{\al }$ and $\hat u^c_{\al }$ with
$\hat d^c_{\al }$, respectively ($\al=1,2,3$).}
\beq
\hat q(6_L)=\hat q(3, 2)_L~,~~~~\hat q^c(6_R)=\hat q^c(3, 2)_R~.
\la{6-dec}
\eeq
In these decompositions, $\hat q$ and $\hat q^c$ can be written as matrices,
\beq
\begin{array}{ccc}
 & {\begin{array}{ccc}
 &~~  ~~  {~}_{\leftarrow ~SU(3)_{fL} ~\rightarrow }&\\
\end{array}}\\
&{\! \hat q= \left(\begin{array}{ccc}

 \hat u& ~~\hat c  & ~~\hat t
\\
 \hat d &~~ \hat s & ~~\hat b
\end{array}\right)}
\end{array}  \!\!\!\!\!\!\!
\begin{array}{c}
 \\ {~}_{\uparrow } \\ \!  {~}_{SU(2)_L}\! \\ \vspace{0.3cm}{~}_{\downarrow } ~
 \end{array}~~~,~~~~~~
 \begin{array}{ccc}
 & {\begin{array}{ccc}
 &~~  ~~  {~}_{\leftarrow ~SU(3)_{fR} ~\rightarrow }&\\
\end{array}}\\
&{\! \hat q^c= \left(\begin{array}{ccc}

 \hat u^c& ~~\hat c^c  & ~~\hat t^c
\\
 \hat d^c &~~ \hat s^c & ~~\hat b^c
\end{array}\right)}
\end{array}  \!\!\!\!\!\!\!
\begin{array}{c}
 \\ {~}_{\uparrow } \\ \!  {~}_{SU(2)_R}\! \\ \vspace{0.3cm}{~}_{\downarrow } ~
 \end{array}~,~
\label{nu-matrix}
\eeq
where schematically actions of $SU(3)$ and $SU(2)$ rotations are depicted.
Therefore,  transformation properties under the chiral group
\beq
G_f^{(3,2)}=SU(3)_{fL}\otimes SU(2)_L \tm SU(3)_{fR}\otimes SU(2)_R\tm U(1)_{B'}
\la{Gf32}
\eeq
are:
\beq
G_f^{(3,2)}~:~~~\hat q\sim \l 3_{fL}, ~2_L,~ 1,~ 1,~ \fr{1}{3}\r ~,~~~~~~
\hat q^c\sim \l 1, ~1,~ 3_{fR},~ 2_R,~ -\fr{1}{3}\r ~.~
\la{trans-Gf32}
\eeq
Relevant anomalies that donot vanish are
$$
A\l [SU(3)_{fL}]^2\!\!\cdot U(1)_{B'}\!\r \!=\!-A\l [SU(3)_{fR}]^2\!\!\cdot U(1)_{B'}\!\r =\! 1~,
$$
\beq
A\l [SU(2)_L]^2\!\!\cdot U(1)_{B'}\!\r \!=\!-A\l [SU(2)_R]^2\!\!\cdot U(1)_{B'}\!\r =\! \fr{3}{2}~.
\la{anom-Gf32}
\eeq
The anomaly matching condition can be satisfied with the spontaneous breaking of the symmetries $SU(3)_{fL}$ and
$SU(3)_{fR}$ down to   $SU(2)_{fL}$ and $SU(2)_{fR}$, respectively. [This happens by condensates (\ref{condens-SM})
discussed above.]
Thus, the chiral symmetry $G_f^{(3,2)}$ is broken down to
$G_f^{(2,2)}$, where
\beq
G_f^{(2,2)}=SU(2)_{fL}\otimes SU(2)_L \tm SU(2)_{fR}\otimes SU(2)_R\tm U(1)_{B'}~.
\la{Gf22}
\eeq
This breaking is realized, for instance,  by the condensates $\lan \hat u_3\hat d_3T_{H'}^\dag \ran $ and $\lan \hat u^c_3\hat d^c_3T_{H'} \ran $.
Note that with $SU(3)_{fL}\to SU(2)_{fL}$ and $SU(3)_{fR}\to SU(2)_{fR}$ we will have decompositions
$3_{fL}=2_{fL}+1_{fL}$ and $3_{fR}=2_{fR}+1_{fR}$. At the composite level, the spin-1/2  and $SU(3)'$ singlet
combinations
$(\hat q \hat q)\hat q$ and $(\hat q^c \hat q^c)\hat q^c$ picked up as $[2'_{fL}+1'_{fL}]$ from $[2_{fL}+1_{fL}]^3$ and
$[2'_{fR}+1'_{fR}]$ from $[2_{fR}+1_{fR}]^3$.
Thus, transformations of $(\hat q \hat q)\hat q$ and $(\hat q^c \hat q^c)\hat q^c$ composites
under $G_f^{(2,2)}$ are\footnote{Under combination $(\hat q \hat q)\hat q$ (suppressed gauge/chiral indices), we mean
$\ep^{a'b'c'}\!\ep_{ij}(\hat q_{a'i} \hat q_{b'j})\hat q_{c'k}$, where $a', b', c'=1,2,3$ are $SU(3)'$ indices and
$i,j,k=1,2$ stand for $SU(2)_L/SU(2)_w$ indices. Similar is applied to the combination $(\hat q^c \hat q^c)\hat q^c$.}
\beq
G_f^{(2,2)}~:~~~(\hat q \hat q)\hat q\sim \l [2_{fL}+1_{fL}], ~2_L,~ 1,~ 1,~ 1\r ~,~~~~~~
(\hat q^c \hat q^c)\hat q^c\sim \l 1, ~1,~ [2_{fR}+1_{fR}],~ 2_R,~ -1\r ~.
\la{comp-trans-Gf22}
\eeq
These representations will have anomalies that precisely match with those given in Eq. (\ref{anom-Gf32}).
Thus, we have three families of  $l_0, e^c_0, \nu^c_0$ composite states represented in Eq. (\ref{comp-st}), with
transformation properties under $G_{SM}$ given in Eq. (\ref{comp-transl}).

\renewcommand{\theequation}{B.\arabic{equation}}\setcounter{equation}{0}

\section{RG equations and $b$ factors}
\la{rg}

In this appendix we discuss details of gauge coupling unification and present one- and two-loop RG
coefficients at each relevant energy scale. At the end we  calculate short-range renormalization factors
$A_S^l$ and $A_S^{e^c}$ for baryon number violating $d=6$ operators.

The two-loop RG equation, for  gauge coupling $\al_i$,  has the form \cite{jonesRG}
\beq
\fr{d}{d\ln \mu}\al_i^{-1}=-\fr{b_i}{2\pi}-\fr{1}{8\pi^2}\sum_jb_{ij}\al_j+\fr{1}{32\pi^3}\sum_fa_i^{f}\lam_f^2 ,
\la{gen-2loopRG}
\eeq
where $b_i$ and $b_{ij}$ account for one- and two-loop gauge contributions, respectively, and $c_i^{f}$ represents the two-loop correction via
 Yukawa coupling $\lam_f$. For consistency, it is enough to consider the Yukawa coupling RG at the one-loop approximation:
 \beq
 16\pi^2\fr{d}{d\ln \mu}\lam_f=c_f\lam_f^3+\lam_f(\sum_{f'}d_f^{f'}\lam_{f'}^2-4\pi \sum_ic_f^i\al_i)~.
 \la{Y-RG}
 \eeq
  RG factors can be calculated using general formulas \cite{jonesRG}.
 Since at different energy scales different states appear, these factors also change with energy. For instance, at scale
 $\mu $, the $b_i$ and $b_{ij}$ can be written as $b_i(\mu )=\sum_a \te (\mu -M_a)b_i^a$ and $b_{ij}(\mu )=\sum_a \te (\mu -M_a)b_{ij}^a$,
 where $a$ stands for the state with mass $M_a$ and step function $\te(x)=0$ for $x\leq 0$, and $\te(x)=1$ for $x> 0$.

Integration of Eq. (\ref{gen-2loopRG}), in energy interval $\mu_1-\mu_2$,  gives
\beq
\al_i^{-1}(\mu_2)=\al_i^{-1}(\mu_1)-\fr{b_i^{\mu_1\mu_2}}{2\pi}\ln \fr{\mu_2}{\mu_1}~,
\la{sol-gen-2loopRG}
\eeq
where an effective $b_i^{\mu_1\mu_2}$ factor is given by
\beq
b_i^{\mu_1\mu_2}=\!\!\l \!\sum_a\te(\mu_2-M_a)b_i^a\ln \fr{\mu_2}{M_a}+
\fr{1}{4\pi}\sum_a\!\!\int_{\mu_1}^{\mu_2}\!\!\!\te(\mu-M_a)b_{ij}^a\al_jd\ln \mu -
\fr{1}{8\pi^2}\int_{\mu_1}^{\mu_2}\!\!\!c_i^f\lam_f^2d\ln \mu \r \!\fr{1}{\ln \fr{\mu_2}{\mu_1}} ~.
\la{gen-eff-b}
\eeq
The second and third terms in Eq. (\ref{gen-eff-b}) can be evaluated iteratively \cite{Langacker:1992rq}.
Although Eq. in (\ref{gen-2loopRG}) can be solved numerically
(which we do perform for obtaining final results), expressions (\ref{sol-gen-2loopRG}) and (\ref{gen-eff-b}) are useful for
understanding how unification works.

In the energy interval $M_Z-\La'$, we have just SM, while between $\La'$ and $M_I$ scales, we have $G_{SM}\tm SU(3)'$ gauge interactions plus
additional states.
Applying Eq. (\ref{sol-gen-2loopRG}) for the couplings $\al_Y, \al_w, \al_c$, and $\al_{3'}$, we will have
$$
\al_i^{-1}(M_I)=\al_i^{-1}(M_Z)-\fr{b_i^{ZI}}{2\pi }\ln \fr{M_I}{M_Z}~,~~~~i=Y, w, c ~,
$$
\beq
\al_{3'}^{-1}(M_I)=\al_{3'}^{-1}(\La')-\fr{b_{3'}^{\La'I}}{2\pi }\ln \fr{M_I}{\La'}~,
\la{sm3prRG-sol}
\eeq
where $b_i^{ZI}, b_{3'}^{\La'I}$ can be calculated via Eq. (\ref{gen-eff-b}) having appropriate RG factors.

Above the scale $M_I$, we have gauge interactions $G_{321}$ going all the way up to the GUT scale.  The ${G_{321}}'$ gauge symmetry
appears between scales $M_I$ and ${M_I}'$, while $SU(5)'$ appears above the ${M_I}'$ scale. Therefore,
we will have
$$
\al_i^{-1}(M_G)=\al_i^{-1}(M_I)-\fr{b_i^{IG}}{2\pi }\ln \fr{M_G}{M_I}~,~~~~i=1, 2, 3 ~,
$$
$$
\al_{i'}^{-1}({M_I}')=\al_{i'}^{-1}(M_I)-\fr{b_{i'}^{II'}}{2\pi }\ln \fr{{M_I}'}{M_I}~,~~~~i'=1', 2', 3' ~,
$$
\beq
\al_{5'}^{-1}(M_G)=\al_{5'}^{-1}({M_I}')-\fr{b_{5'}^{I'G}}{2\pi }\ln \fr{M_G}{{M_I}'}~.
\la{sols-aboveMI}
\eeq
From Eqs. (\ref{sm3prRG-sol}) and (\ref{sols-aboveMI}) and taking into account the boundary conditions (\ref{bound-Yw})-(\ref{unif-MG}),
 we arrive at relations given in Eq. (\ref{calc-unif}). The four equations in  (\ref{calc-unif}) allow us to determine $M_I, {M_I}', M_G$ and $\al_G$,
in terms of other input mass scales. The latter must be selected in such a way as to get successful unification. This has been done
numerically, and results are given in Table \ref{t:tab1}, Eq. (\ref{output}), and Fig. \ref{fig3}.

Now we present all RG $b$ factors needed for writing down RG equations.
In the energy interval $\mu =M_Z-\La'$, the RG factors are just those of the SM:
\beq
\mu =M_Z - \La'~:~~~ b_i=\l \fr{41}{10}, -\fr{19}{6}, -7\r ~,
b_{ij}=\left(
  \begin{array}{ccc}
    \fr{199}{50} & \fr{27}{10} & \fr{44}{5} \\
    \fr{9}{10} & \fr{35}{6} & 12 \\
    \fr{11}{10} & \fr{9}{2} & -26 \\
  \end{array}
\right)~,~
(i=Y, w, c)~.
\la{b-MZ-Lam1}
\eeq

In the energy interval $\La' - M_I$, we have the symmetry $SU(3)_c\tm SU(2)_w\tm U(1)_Y\tm SU(3)'$.
Also, instead of composite leptons, we have three families of
 $SU(3)'$ triplets  $\hat q, \hat u^c, \hat d^c$, and vectorlike states $(l, \hat l)_{\al}$ and $(e^c, \hat e^c)_{\al}$ ($\al =1,2,3$)
 with masses $M_{\hat ll}^{(\al)}$ and $M_{e^c\hat e^c}^{(\al)}$, respectively. Moreover, some fragments of $\Phi(5, \bar 5)$ [see Eq. (\ref{Phi55-dec})]
 and $\Si'_{8'}$ (of $\Si'$) can appear below $M_I$.
Thus, the corresponding $b$ factors in this energy interval are given by
$$
\mu =\La' - M_I :
$$
$$
 b_Y=\fr{9}{2}\!+\!\fr{1}{15}\te \!\l \mu-M_{T_{H'}} \!\r
\!+\!\fr{2}{5}\sum_{\al=1}^3\!\te \!\l \mu \!-\!M_{\hat ll}^{(\al)}\!\r
\!+\!\fr{4}{5}\sum_{\al=1}^3\!\te \!\l \mu \!-\!M_{e^c\hat e^c}^{(\al)}\!\r +\fr{5}{6}\!\te \!\l \mu \!-\!M_{DT'}\!\r
 +\fr{5}{6}\!\te \!\l \mu \!-\!M_{TD'}\!\r
 $$
 $$
b_w=-\fr{7}{6}\!+\!\fr{2}{3}\sum_{\al=1}^3\!\te \!\l \mu \!-\!M_{\hat ll}^{(\al)}\!\r +\fr{1}{2}\!\te \!\l \mu \!-\!M_{DT'}\!\r
+\fr{1}{2}\!\te \!\l \mu \!-\!M_{TD'}\!\r,
$$
$$
b_c=-7+\fr{1}{3}\!\te \!\l \mu \!-\!M_{TD'}\!\r +\fr{1}{2}\!\te \!\l \mu \!-\!M_{TT'}\!\r,~~
$$
\beq
 b_{3'}=-7\!+\!\fr{1}{6}\te \!\l \mu \!-\!M_{T_{H'}}\!\r +\fr{1}{3}\!\te \!\l \mu \!-\!M_{DT'}\!\r +\fr{1}{2}\!\te \!\l \mu \!-\!M_{TT'}\!\r
+\fr{1}{2}\!\te \!\l \mu \!-\!M_{8'}\!\r ~,
\la{b-Lam1-MI}
\eeq
$$
\mu =\La' - M_I :~~~b_{ij}=
\l \!\!
  \begin{array}{cccc}
    \fr{13709}{50} & \fr{9}{5} & \fr{44}{5} & \fr{44}{5} \\
    \fr{3}{5} & \fr{91}{3} & 12 & 12 \\
     \fr{11}{10}& \fr{9}{2} & -26 & 0 \\
    \fr{11}{10} & \fr{9}{2} & 0 & -26 \\
  \end{array} \!\!\r
\!+\sum_a \te \!\l \mu \!-\!M_a\!\r b_{ij}^a~,~~~~~(i,j=Y, w, c, 3')~~~{\rm with:}
$$
$$
b_{ij}^{T_{H'}}\!\!=\!\!\l \!\!
  \begin{array}{cccc}
    \fr{4}{75} & 0 & 0 & \fr{16}{15} \\
    0 & 0 & 0 & 0 \\
    0 & 0 & 0 & 0 \\
    \fr{2}{15} & 0 & 0 & \fr{11}{3} \\
  \end{array} \!\!\r \!\!,~~
  b_{ij}^{DT'}\!\!=\!\!\l \!\!
  \begin{array}{cccc}
    \fr{25}{6} & \fr{15}{2} & 0 & \fr{40}{3} \\
    \fr{5}{2} & \fr{13}{2} & 0 & 8 \\
    0 & 0 & 0 & 0 \\
    \fr{5}{3} & 3 & 0 & \fr{22}{3} \\
  \end{array} \!\!\r \!\!,~~
  b_{ij}^{TT'}\!\!=\!\!\l \!\!
  \begin{array}{cccc}
    0 & 0 & 0 & 0 \\
    0& 0 & 0 & 0 \\
    0 & 0 & 11 & 8 \\
    0 & 0 & 8 & 11 \\
  \end{array} \!\!\r \!\!,~~
  b_{ij}^{TD'}\!\!=\!\!\l \!\!
  \begin{array}{cccc}
    \fr{25}{6} & \fr{15}{2} & \fr{40}{3} & 0 \\
    \fr{5}{2}& \fr{13}{2} & 8 & 0 \\
    \fr{5}{2} & 3 & \fr{22}{3} & 0 \\
    0 & 0 & 0 & 0 \\
  \end{array} \!\!\r \!\!,
$$
\beq
b_{ij}^{(l, \hat l)_{\al}}\!\!=\!\!\l \!\!
  \begin{array}{cccc}
    \fr{9}{50} & \fr{9}{10} & 0 & 0 \\
    \fr{3}{10} & \fr{49}{6} & 0 & 0 \\
    0 & 0 & 0 & 0 \\
    0 & 0 & 0 & 0 \\
  \end{array} \!\!\r \!\!,~~
  b_{ij}^{(e^c, \hat e^c)_{\al}}\!\!=\!\!{\rm Diag}\!\l \! \fr{36}{25}, 0, 0, 0\! \r \! ,~~
   b_{ij}^{\Si'_{8'}}\!\!=\!\!{\rm Diag}\!\l \! 0, 0, 0, 21\! \r \! ~.
\la{2loop-b-Lam1-MI}
\eeq

Between the scales $M_I$ and ${M_I}'$, the symmetry is $G_{321}\tm {G_{321}}'$, and all matter states are massless. Also, above
the scale $M_I$,  we should include the states $T_{H'}$ and $\Phi_{DD'}$ as massless and remaining fragments above their mass thresholds.
Since $G_{321}$ goes all the way up to the $M_G$, its one-loop b factors can be determined in the interval $M_I-M_G$ and are given by
\begin{eqnarray}
\mu =M_I - M_G :&~ b_1\!&\!=\fr{43}{10}\!+\!\fr{3}{10}\te \!\l \mu-M_{DT'} \!\r
\!+\! \fr{1}{5}\te \!\l \mu-M_{TT'} \!\r \!+\! \fr{2}{15}\te \!\l \mu-M_{TD'} \!\r ,\nonumber \\
~ & ~ b_2&\!=-\fr{17}{6}\!+\!\fr{1}{2}\te \!\l \mu-M_{DT'} \!\r ,\nonumber \\
~ & ~ b_3&\!=-7\!+\!\fr{1}{2}\te \!\l \mu-M_{TT'} \!\r  \!+\!\fr{1}{3}\te \!\l \mu-M_{TD'} \!\r .
\la{b-MI-MG}
\end{eqnarray}
The gauge group  ${G_{321}}'$ appears in the interval $M_I-{M_I}'$, and corresponding one-loop $b$ factors are
\begin{eqnarray}
\mu =M_I - {M_I}' :&~ b_{1'}\!&\!=\fr{64}{15}\!+\!\fr{1}{10}\te \!\l \mu-M_{D'} \!\r
\!+\!\fr{2}{15}\te \!\l \mu-M_{DT'} \!\r
\!+\! \fr{1}{5}\te \!\l \mu-M_{TT'}\!\r  \nonumber \\
&& +\! \fr{3}{10}\te \!\l \mu-M_{TD'}\! \r \!-\! \fr{55}{3}\te \!\l \mu-M_{X'} \!\r ,\nonumber \\
~ & ~ b_{2'}&\!=-3\!+\!\fr{1}{6}\te \!\l \mu-M_{D'} \!\r
\!+\!\fr{1}{2}\te \!\l \mu-M_{TD'} \!\r \!-\! 11\te \!\l \mu-M_{X'}\!\r ,\nonumber \\
~ & ~ b_{3'}&\!=-\fr{41}{6}\!+\!\fr{1}{2}\te \!\l \mu-M_{TT'} \!\r  \!+\!\fr{1}{3}\te \!\l \mu-M_{DT'} \!\r ,
\la{b1-MI-MI1}
\end{eqnarray}
where terms with $\te \!\l \mu-M_{X'} \!\r$ account for the threshold of $(X',Y')$ gauge bosons of $SU(5)'$, in case their masses $M_{X'}$
lie slightly below the ${M_I}'$ scale. We will take this effect into account at 1-loop level.
The two-loop $b_{ij}$ factors of $G_{321}\tm {G_{321}}'$ form $6\tm 6$ matrices and are determined in the interval $M_I - {M_I}'$:
$$
\mu =M_I - {M_I}':~~b_{ij}=(b^{f}\!+\!b^{h}\!+\!b^g\!+\!b^{T_{H'}}\!+\!b^{DD'})_{ij}
+\sum_a \te \!\l \mu \!-\!M_a\!\r b_{ij}^a~,~~~~~(i,j=1, 2, 3, 1', 2', 3')
$$
$$
{\rm with:}~~
b_{ij}^f=3\l \!\!
  \begin{array}{cccccc}
    \fr{19}{15} & \fr{3}{5} & \fr{44}{15} & 0 & 0 & 0 \\
    \fr{1}{5} & \fr{49}{3} & 4 & 0 & 0 & 0 \\
    \fr{11}{30} & \fr{3}{2} & \fr{76}{3} & 0 & 0 & 0 \\
    0 & 0 & 0 & \fr{19}{15} & \fr{3}{5} & \fr{44}{15} \\
    0 & 0 & 0 & \fr{1}{5} & \fr{49}{3} & 4 \\
    0 & 0 & 0 & \fr{11}{30} & \fr{3}{2} & \fr{76}{3} \\
  \end{array} \!\!\r,~~
b_{ij}^h=\l \!\!
  \begin{array}{cccccc}
    \fr{9}{50} & \fr{9}{10} & 0 & 0 & 0 & 0 \\
    \fr{3}{10} & \fr{13}{6} & 0 & 0 & 0 & 0 \\
    0 & 0 & 0 & 0 & 0 & 0 \\
    0 & 0 & 0 & 0 & 0 & 0 \\
    0 & 0 & 0 & 0 & 0 & 0 \\
    0 & 0 & 0 &0 &0 & 0 \\
  \end{array} \!\!\r \!,
$$
$$
b_{ij}^{T_{H'}}\!=\!\!\l \!\!
  \begin{array}{cccccc}
    0 & 0 & 0 & 0 & 0 & 0 \\
   0 & 0 & 0 & 0 & 0 & 0 \\
    0 & 0 & 0 & 0 & 0 & 0 \\
    0 & 0 & 0 & \fr{4}{75} & 0 & \fr{16}{15}  \\
    0 & 0 & 0 & 0 & 0 & 0 \\
    0 & 0 & 0 &\fr{2}{15}  &0 & \fr{11}{3}  \\
  \end{array} \!\!\r \!,~~
  b_{ij}^{DD'}\!=\!\!\l \!\!
  \begin{array}{cccccc}
    \fr{9}{25} & \fr{9}{5} & 0 & \fr{9}{25} & \fr{9}{5} & 0 \\
   \fr{3}{5} & \fr{13}{3} & 0 & \fr{3}{5} & 3 & 0 \\
    0 & 0 & 0 & 0 & 0 & 0 \\
    \fr{9}{25} & \fr{9}{5} & 0 & \fr{9}{25} & \fr{9}{5} & 0  \\
    \fr{3}{5} & 3 & 0 & \fr{3}{5} & \fr{13}{3} & 0 \\
    0 & 0 & 0 &0 &0 & 0 \\
  \end{array} \!\!\r \!,~~
  b_{ij}^{DT'}\!=\!\!\l \!\!
  \begin{array}{cccccc}
    \fr{27}{50} & \fr{27}{10} & 0 & \fr{6}{25} & 0 & \fr{24}{5} \\
   \fr{9}{10} & \fr{13}{2} & 0 & \fr{2}{5} & 0 & 8 \\
    0 & 0 & 0 & 0 & 0 & 0 \\
    \fr{6}{25} & \fr{6}{5} & 0 & \fr{8}{75} & 0 & \fr{32}{15}  \\
    0 & 0 & 0 & 0 & 0 & 0 \\
    \fr{9}{15} & 3 & 0 &\fr{4}{15} &0 & \fr{22}{3} \\
  \end{array} \!\!\r \!,
  $$
  $$
  b_{ij}^{TT'}\!=\!\!\l \!\!
  \begin{array}{cccccc}
    \fr{4}{25} & 0 & \fr{16}{5} & \fr{4}{25} & 0 & \fr{16}{5} \\
   0 & 0 & 0 & 0 & 0 & 0 \\
    \fr{2}{5} & 0 & 11 & \fr{2}{5} & 0 & 8 \\
    \fr{4}{25} & 0 & \fr{16}{5} & \fr{4}{25} & 0 & \fr{16}{5}  \\
    0 & 0 & 0 & 0 & 0 & 0 \\
    \fr{2}{5} & 0 & 8 &\fr{2}{5} &0 & 11 \\
  \end{array} \!\!\r \!,~~
b_{ij}^{TD'}\!=\!\!\l \!\!
  \begin{array}{cccccc}
    \fr{8}{75} & 0 & \fr{32}{15} & \fr{6}{25} & \fr{6}{5}  & 0 \\
   0 & 0 & 0 & 0 & 0 & 0 \\
    \fr{4}{15} & 0 & \fr{22}{3} & \fr{3}{5} & 3 & 0 \\
    \fr{6}{25} & 0 & \fr{24}{5} & \fr{27}{50} & \fr{27}{10} & 0  \\
    \fr{2}{5} & 0 & 8 & \fr{9}{10} & \fr{13}{2} & 0 \\
    0 & 0 & 0 &0 &0 & 0 \\
  \end{array} \!\!\r \!,~
  b_{ij}^{D'}\!=\!\!\l \!\!
  \begin{array}{cccccc}
    0 & 0 &0 & 0 & 0  & 0 \\
   0 & 0 & 0 & 0 & 0 & 0 \\
   0 & 0 & 0 & 0 & 0 & 0 \\
   0 & 0 & 0 & \fr{9}{50} & \fr{9}{10} & 0  \\
   0 & 0 & 0 & \fr{3}{10} & \fr{13}{6} & 0 \\
    0 & 0 & 0 &0 &0 & 0 \\
  \end{array} \!\!\r \!,
  $$
  \beq
b_{ij}^g=\!{\rm Diag}\! \l \!0, -\fr{136}{3}, -102, 0, -\fr{136}{3}, -102\!\r \!,~~
b_{ij}^{\Si'_{8'}}=\!{\rm Diag}\! \l \!0, 0, 0, 0, 0, 21\!\r \!.
  \la{2loop-b-MI-MI1}
\eeq
In this $M_I-{M_I}'$ energy interval, we have two Abelian factors $U(1)$ and $U(1)'$ and states $\Phi_i$ (the fragments of
$\Phi$) charged under both gauge symmetries. Because of this, the gauge kinetic mixing will be induced \cite{Holdom:1985ag}, \cite{Babu:1996vt}.
Parametrizing the latter as $-\fr{\sin \chi }{2}F_1^{\mu \nu}F_{1'\mu \nu}$, and bringing whole gauge kinetic part to the
canonical form, one can obtain $\Phi_i$'s covariant derivative as \cite{Babu:1996vt}:
$[\pl^{\mu}+\fr{i}{2}g_1Q_iA_1^{\mu }+\fr{i}{2}(\bar g_{1'}{Q_i}'+g_{11'}Q_i)A_{1'}^{\mu }]\Phi_i$. In this basis $Q_i$ charges are unshifted, and
$g_1$ and its RG are unchanged. On the other hand, $\bar g_{1'}=g_{1'}/\cos \chi $ and $g_{11'}=-g_1 \tan \chi $. Introducing the ratio
$\de =g_{11'}/\bar g_{1'}$, the RGs for $\bar \al_{1'}$ and $\de $ will be \cite{Babu:1996vt}
\beq
\fr{d}{d\ln \mu }(\bar \al_{1'})^{-1}=\cdots -\fr{b_1}{2\pi}\de^2-\fr{B_{11'}}{\pi }\de ~,~~~~~~
\fr{d}{d\ln \mu } \de =\fr{b_1}{2\pi}\al_1\de+\fr{B_{11'}}{8\pi^2}~,
\la{shifterRG}
\eeq
where $"\dots "$ denote standard one- and two-loop contributions [with form of Eq. (\ref{gen-2loopRG})] and
$B_{11'}=\sum_iQ_i{Q_i}'$ is given by
\beq
B_{11'}=\fr{1}{5}\left [ \te (\mu-M_{DT'})-\te (\mu-M_{DD'})-\te (\mu-M_{TT'})+\te (\mu-M_{TD'})\right ] ~.
\la{mixB11}
\eeq
Because of the mass splitting  between $\Phi $'s fragments, $B_{11'}\neq 0$ in the interval $M_I-M_{TD'}$, and therefore $\de \neq 0$;
i.e., the kinetic mixing is generated. This causes the shift $\al_{1'}^{-1}\to \al_{1'}^{-1}+{\cal O}(\de )$. However, as it turns out, within our
model this effect is negligible. We have taken these into account upon numerical studies and got $\de(M_I)\simeq 9.5\cdot 10^{-3}$,
$\sin \chi (M_I)\simeq -2\cdot 10^{-2}$, causing the change of $\al_{1'}^{-1}(M_I)$ by $0.01\%$. This has no practical impact
on the matching conditions of Eq. (\ref{bound-Yw}), does not affect the picture of gauge coupling unification and therefore can be safely ignored.

Since at and above the scale ${M_I}'$ the  ${G_{321}}'$ is embedded in $SU(5)'$, we will deal with b factors of $G_{321}\tm SU(5)'$ symmetry,
and one-loop b factors of $G_{321}$ are given in Eq. (\ref{b-MI-MG}). At energies corresponding to unbroken $SU(5)'$,
 the fragments $(\Phi_{DD'}, \Phi_{DT'})$ form the unified $(2,\bar 5)\equiv \Phi_{D\bar 5'}$-plet of $G_{321}\tm SU(5)'$.
 Similarly, $(T_{H'}, D')\subset H'$.
 Above the scale ${M_I}'$, these states (together with all fragments of the $\Si'$-plet) should be included as massless states.
Thus, the one-loop  $b$ factor of $SU(5)'$ is given as
\beq
\mu ={M_I}' - M_G~ :~~~b_{5'}=-13\!+\!\fr{1}{2}\te \!\l \mu-M_{T\bar 5'} \!\r ,
\la{b51-MI1-MG}
\eeq
where $M_{T\bar 5'}={\rm max}(M_{TT'}, M_{TD'})$ denotes the mass of the $(3,\bar 5)$-plet, which includes $\Phi_{TT'}$ and $\Phi_{TD'}$ states:
$(\Phi_{TT'}, \Phi_{TD'})\subset \Phi_{T\bar 5'}$.
The two-loop $b_{ij}$ factors, above the scale ${M_I}'$, form $4\tm 4$ matrices and are
$$
\mu ={M_I}' - M_G:~~b_{ij}=(b^{f}\!+\!b^{h}\!+\!b^g\!+\!b^{H'}\!+\!b^{\Si'}\!+\!b^{D\bar 5'})_{ij}
+\te \!\l \mu \!-\!M_{T\bar 5'}\!\r b_{ij}^{T\bar 5'}~,~~~~~(i,j=1, 2, 3, 5')
$$
$$
{\rm with}:~
b^{f}_{ij}=3\l \!\!\begin{array}{cccc}
    \fr{19}{15} & \fr{3}{5} & \fr{44}{15} & 0 \\
    \fr{1}{5} & \fr{49}{3} & 4 & 0 \\
    \fr{11}{30} & \fr{3}{2} & \fr{76}{3} & 0 \\
    0 & 0 & 0 & \fr{698}{15} \\
  \end{array} \!\!\r \!,~~
  b^{h}_{ij}=\l \!\!\begin{array}{cccc}
    \fr{9}{50} & \fr{9}{10} & 0 & 0 \\
    \fr{3}{10} & \fr{13}{6} & 0 & 0 \\
    0 & 0 & 0 & 0 \\
    0 & 0 & 0 & 0 \\
  \end{array} \!\!\r \!,~~
  b_{ij}^g=\!{\rm Diag}\! \l \!0, -\fr{136}{3}, -102, -\fr{850}{3}\!\r \!,~~
$$
\beq
b^{H'}_{ij}=\fr{97}{15}\de_{i5'}\de_{j5'} ,~~b^{\Si'}_{ij}=\fr{175}{3}\de_{i5'}\de_{j5'} ,~~
b^{D\bar 5'}_{ij}=\l \!\!\begin{array}{cccc}
    \fr{9}{10} & \fr{9}{2} & 0 & \fr{72}{5} \\
    \fr{3}{2} & \fr{65}{6} & 0 & 24 \\
    0 & 0 & 0 & 0 \\
    \fr{3}{5}  & 3 & 0 & \fr{194}{15}  \\
  \end{array} \!\!\r \!,~~
  b^{T\bar 5'}_{ij}=\l \!\!\begin{array}{cccc}
    \fr{4}{15} & 0 & \fr{16}{3}  & \fr{48}{5}  \\
    0 & 0 & 0 & 0 \\
    \fr{2}{3}  & 0 & \fr{55}{3}  & 24 \\
    \fr{2}{5}  & 0 & 8 & \fr{97}{5}  \\
  \end{array} \!\!\r \!.
  \la{2loop-b-MI1-MG}
\eeq

As far as the Yukawa coupling involving RG factors, $a_i^f, c_f, d_f^{f'}$, and $c_f^i$ [see Eqs. (\ref{gen-2loopRG}) and (\ref{Y-RG})],
are concerned, within our model only top and "mirror-top" Yukawa couplings are large.
All other Yukawa interactions are small and can be ignored. Thus, the Yukawa terms $\lam_tq_3t^ch$,
$(\lam_{\hat t\hat b}\hat t\hat b+\lam_{\hat t^c\hat{\tau}^c}\hat t^c\hat{\tau}^c)T_{H'}$, and $\lam_{\hat t}\hat q_3\hat t^cD'$
are relevant. All these four couplings unify at $M_G$ due to gauge symmetry and $D_2$ parity.
For the top Yukawa involved RG factors, in the  energy interval $M_Z-M_I$, we have
\beq
a_i^t=\l \fr{17}{10}, \fr{3}{2}, 2\r ,~~~c_t^i=\l \fr{17}{20}, \fr{9}{4}, 8\r ,~~~(i=Y, w, c),~~~c_t=\fr{9}{2}, ~~~d_t^{f'}=0~.
\la{topRG-factors}
\eeq
In energy interval $M_I-M_G$, with replacement of the indices $(Y, w, c)\to (1, 2, 3)$, the corresponding RG factors will be the same.
Since the mass of  the state $D'$  is $\sim {M_I}'$, the  RG with $\lam_{\hat t}$ will be relevant above the scale ${M_I}'$.
Within our model, $M_{T_{H'}}\sim \La'$, and in the RG, the couplings $\lam_{\hat t\hat b}$ and $\lam_{\hat t^c\hat{\tau}^c}$ will be relevant above
the scale  $\La'$. Between
the scales $\La'$ and $M_I$, the mirror matter has EW and $SU(3)'$ interactions. Therefore, we have
$$
\mu=\La'-M_I~:~~~(a_Y, a_w, a_{3'})^{\hat t\hat b}=\l \fr{1}{15}, 2, \fr{4}{3}\r ,~~~
(a_Y, a_w, a_{3'})^{\hat t^c\hat{\tau}^c}\!=\te (\mu\!-\!M_{e^c\hat{e}^c}^{(3)})\l \fr{13}{15}, 0, \fr{1}{3}\r ,
$$
$$
\hs{1cm}(c^Y, c^w, c^{3'})_{\hat t\hat b}=\l \fr{1}{10}, \fr{9}{2}, 8\r ,~~~
(c^Y, c^w, c^{3'})_{\hat t^c\hat{\tau}^c}\!=\te (\mu\!-\!M_{e^c\hat{e}^c}^{(3)})\l \fr{13}{5}, 0, 4\r ,
$$
\beq
\hs{3cm} c_{\hat t\hat b}=4,~~ d_{\hat t\hat b}^{\hat t^c\hat{\tau}^c}=\te (\mu\!-\!M_{e^c\hat{e}^c}^{(3)}),~~
c_{\hat t^c\hat{\tau}^c}=3\te (\mu\!-\!M_{e^c\hat{e}^c}^{(3)}),~~d^{\hat t\hat b}_{\hat t^c\hat{\tau}^c}=2\te (\mu\!-\!M_{e^c\hat{e}^c}^{(3)}).
\la{mirRG-factors}
\eeq
Between $M_I$ and ${M_I}'$ scales, with replacements $(Y, w)\to (1', 2')$, the corresponding factors will be the same.
At and above the scale $M_I$, the ${G_{321}}'$ is unified in the $SU(5)'$ group, $D'$ should be included in the RG, and three Yukawas unify
$\lam_{\hat t\hat b}=\lam_{\hat t^c\hat{\tau}^c}=\lam_{\hat t}$. Thus, dealing with $\lam_{\hat t}$, we will have
\beq
\mu={M_I}'-M_G~:~~~~a_{5'}^{\hat t}=\fr{9}{2},~~~~c_{\hat t}=9,~~~~c_{\hat t}^{5'}=\fr{108}{5},~~~~d_{\hat t}^{f'}=0~.
\la{SU51-Yuk-RGfactors}
\eeq

  \subsection{Short-range RG factors for $d=6$ operators}
  \la{d6RG}

  The baryon number violating $d=6$ operators of Eq. (\ref{d6-ops-B-L1}) involve couplings
  ${\cal C}^{(e^c)}$ and ${\cal C}^{(l)}$ respectively. These couplings run, and in nucleon decay amplitudes, the short-range
  RG factors
  \beq
 A_S^{l}=\fr{{\cal C}^{(l)}(M_Z)}{{\cal C}^{(l)}(M_X)} ~, ~~~~~~~~~~~A_S^{e^c}=\fr{{\cal C}^{(e^c)}(M_Z)}{{\cal C}^{(e^c)}(M_X)}
  \la{def-AS}
  \eeq
  emerge. These factors, having SM gauge interactions and states below the GUT scale, were calculated in Ref. \cite{Buras:1977yy}.
  Within our model, calculation can be done similarly. The RG equations for ${\cal C}^{(l)}$ and ${\cal C}^{(e^c)}$,  in one-loop approximation,
  are given by
  \begin{eqnarray}
  \la{RG-for-d6AS}
  4\pi \fr{d}{dt}{\cal C}^{(l)}\!&=&\!\!\!-{\cal C}^{(l)}\!\left [ \te (M_I\!-\!\mu)\!\l \!\fr{23}{20}\al_Y+\fr{9}{4}\al_w\!\r+2\al_c
  \!+\!\te (\mu\!-\!M_I)\!\l \!\fr{23}{20}\al_1+\fr{9}{4}\al_2\!\r \right ] ,\nonumber \\
   4\pi \fr{d}{dt}{\cal C}^{(e^c)}\!&=&\!\!\!-{\cal C}^{(e^c)}\!\left [ \te (M_I\!-\!\mu)\!\l \!\fr{11}{20}\al_Y+\fr{9}{4}\al_w\!\r\!+\!2\al_c
  \!+\!\te (\mu\!-\!M_I)\!\l \!\fr{11}{20}\al_1+\fr{9}{4}\al_2\!\r \right ] .
  \end{eqnarray}
  Having numerical solutions for the gauge couplings,  Eqs.  (\ref{RG-for-d6AS}) can be integrated. Doing so and taking into account Eqs. (\ref{def-AS}),
  within our model we obtain $A_S^l=1.18$ and $A_S^{e^c}=1.17$.

\bibliographystyle{unsrt}

\begin{thebibliography}{99}


\bibitem{Aad:2012tfa}
  G.~Aad {\it et al.}  [ATLAS Collaboration],
  Phys.\ Lett.\ B {\bf 716}, 1 (2012); 
  S.~Chatrchyan {\it et al.}  [CMS Collaboration],
    Phys.\ Lett.\ B {\bf 716}, 30 (2012).

\bibitem{Buttazzo:2013uya}
  D.~Buttazzo, G.~Degrassi, P.~P.~Giardino, G.~F.~Giudice, F.~Sala, A.~Salvio and A.~Strumia,
  JHEP {\bf 1312}, 089 (2013).

\bibitem{Sher:1988mj}
  M.~Sher,
   Phys.\ Rept.\  {\bf 179}, 273 (1989).


\bibitem{nu-data}
  G.~L.~Fogli, E.~Lisi, A.~Marrone, D.~Montanino, A.~Palazzo and A.~M.~Rotunno,
    Phys.\ Rev.\ D {\bf 86}, 013012 (2012);
    T. Schwetz, M. Tortola and J.W.F. Valle,
  New J.\ Phys.\  {\bf 13}, 109401 (2011).

\bibitem{Pati:1974yy}
 J.~C.~Pati and A.~Salam,
 Phys.\ Rev.\ Lett.\  {\bf 31}, 661 (1973);
   Phys.\ Rev.\ D {\bf 10}, 275 (1974).

\bibitem{GG-gut}
H.~Georgi and S.~L.~Glashow,
  Phys.\ Rev.\ Lett.\  {\bf 32}, 438 (1974);
  H.~Georgi, H.~R.~Quinn and S.~Weinberg,
  Phys.\ Rev.\ Lett.\  {\bf 33}, 451 (1974) .

  \bibitem{Fritzsch:1974nn}
H. Georgi, {\it Particles and Fields}, edited by C. Carlson (AIP, New York, 1975);
 H.~Fritzsch and P.~Minkowski,
   Annals Phys.\  {\bf 93}, 193 (1975).


\bibitem{seesaw}
P.~Minkowski,
  Phys.\ Lett.\ B {\bf 67}, 421 (1977);
  M.~Gell-Mann, P.~Ramond and R.~Slansky,  {\it Supergravity} edited by
  P. van Nieuwenhuizen and D.Z. Freedman (North Holland, Amsterdam, 1979), p. 315;
  T.~Yanagida,
in {\it Proceedings of the Workshop on the Baryon Number of the Universe and Unified Theories, Tsukuba, Japan, 13-14 Feb 1979};
S.~L.~Glashow,
  NATO Adv.\ Study Inst.\ Ser.\ B Phys.\  {\bf 59}, 687 (1979);
R.~N.~Mohapatra and G.~Senjanovic,
  Phys.\ Rev.\ Lett.\  {\bf 44}, 912 (1980) .


  \bibitem{1product-group-GUT}
  A.~Davidson and K.~C.~Wali,
    Phys.\ Rev.\ Lett.\  {\bf 58}, 2623 (1987);
  P.~L.~Cho,
  Phys.\ Rev.\ D {\bf 48}, 5331 (1993); 
  R.~Barbieri, G.~R.~Dvali and A.~Strumia,
    Phys.\ Lett.\ B {\bf 333}, 79 (1994); 
  P.~H.~Frampton and O.~C.~W.~Kong,
    Phys.\ Rev.\ D {\bf 53}, 2293 (1996); 
  R.~N.~Mohapatra,
 Phys.\ Lett.\ B {\bf 379}, 115 (1996); 
  S.~M.~Barr,
Phys.\ Rev.\ D {\bf 55}, 6775 (1997); 
 E.~Witten,
  hep-ph/0201018;
   M.~Dine, Y.~Nir and Y.~Shadmi,
  Phys.\ Rev.\ D {\bf 66}, 115001 (2002); 
  D.~Emmanuel-Costa, E.~T.~Franco and R.~Gonzalez Felipe,
  JHEP {\bf 1108}, 017 (2011). 



\bibitem{Ellis:1979fg}
E.~Witten,
  Phys.\ Lett.\ B {\bf 91}, 81 (1980);
  J.~R.~Ellis, M.~K.~Gaillard,
  Phys.\ Lett.\ B {\bf 88}, 315 (1979);
  H.~Georgi, C.~Jarlskog,
  Phys.\ Lett.\ B {\bf 86}, 297 (1979); For recent discussion and references, see
  K.~S.~Babu, B.~Bajc and Z.~Tavartkiladze,
   Phys.\ Rev.\ D {\bf 86}, 075005 (2012).


  \bibitem{2product-group-GUT}
  G.~C.~Joshi and R.~R.~Volkas,
  Phys.\ Rev.\ D {\bf 45}, 1711 (1992);
  G.~R.~Dvali and Q.~Shafi,
    Phys.\ Lett.\ B {\bf 339}, 241 (1994); 
  N.~Maekawa and Q.~Shafi,
   Prog.\ Theor.\ Phys.\  {\bf 109}, 279 (2003); 
  K.~S.~Babu, E.~Ma and S.~Willenbrock,
  Phys.\ Rev.\ D {\bf 69}, 051301 (2004); 
  K.~S.~Babu, S.~M.~Barr and I.~Gogoladze,
  Phys.\ Lett.\ B {\bf 661}, 124 (2008); 
  B.~Stech,
   Phys.\ Rev.\ D {\bf 86}, 055003 (2012). 

%
%

\bibitem{Pati:1975md}
  J.~C.~Pati, A.~Salam and J.~A.~Strathdee,
  Phys.\ Lett.\ B {\bf 59}, 265 (1975).


\bibitem{Greenberg:1975yg}
O.~W.~Greenberg and C.~A.~Nelson,
   Phys.\ Rev.\ D {\bf 10}, 2567 (1974);
  O.~W.~Greenberg,
   Phys.\ Rev.\ Lett.\  {\bf 35}, 1120 (1975);
   H.~Terazawa,
  Phys.\ Rev.\ D {\bf 22}, 184 (1980).



\bibitem{hooft}
G. 't Hooft,  {\it Recent Development in Gauge Theories}, edited by G. 't Hooft
{\it et al.} (Plenum, New York, 1980);
  Y.~Frishman, A.~Schwimmer, T.~Banks and S.~Yankielowicz,
   Nucl.\ Phys.\ B {\bf 177}, 157 (1981).


\bibitem{Barbieri:1980aq}
  R.~Barbieri, L.~Maiani and R.~Petronzio,
   Phys.\ Lett.\ B {\bf 96}, 63 (1980);
  S.~J.~Brodsky and S.~D.~Drell,
  Phys.\ Rev.\ D {\bf 22}, 2236 (1980).


\bibitem{Dimopoulos:1980hn}
  S.~Dimopoulos, S.~Raby and L.~Susskind,
  Nucl.\ Phys.\ B {\bf 173}, 208 (1980).


\bibitem{Nilles:1981bx}
  H.~P.~Nilles and S.~Raby,
   Nucl.\ Phys.\ B {\bf 189}, 93 (1981).

\bibitem{Greenberg:1980ri}
  O.~W.~Greenberg and J.~Sucher,
   Phys.\ Lett.\ B {\bf 99}, 339 (1981);
  R.~Barbieri, R.~N.~Mohapatra and A.~Masiero,
   Phys.\ Lett.\ B {\bf 105}, 369 (1981).


\bibitem{comp-no-anom-cond}
%
%
H.~Harari and N.~Seiberg,
Phys.\ Lett.\ B {\bf 102}, 263 (1981);
   Phys.\ Lett.\ B {\bf 115}, 450 (1982).

\bibitem{Buchmuller:1983iu}
  W.~Buchmuller, R.~D.~Peccei and T.~Yanagida,
  Nucl.\ Phys.\ B {\bf 227}, 503 (1983).


  \bibitem{Chkareuli:1982rn}
  J.~L.~Chkareuli,
   JETP Lett.\  {\bf 36}, 493 (1983);
   X.~Li and R.~E.~Marshak,
  Nucl.\ Phys.\ B {\bf 268}, 383 (1986);
   R.~R.~Volkas and G.~C.~Joshi,
  Phys.\ Rept.\  {\bf 159}, 303 (1988).


\bibitem{review-comp}
For reviews, see\\
M.~E.~Peskin, {\it Compositeness of Quarks and Leptons,}  eConf C {\bf 810824}, 880 (1981);\\
  S.~Raby, {\it Light Composite Fermions: An Overview,}  LA-UR-82-1416;\\
  L.~Lyons,
  {\it An Introduction to the Possible Substructure of Quarks and Leptons,}
    Prog.\ Part.\ Nucl.\ Phys.\  {\bf 10}, 227 (1983);\\
  R.~N.~Mohapatra,
  {\it Unification And Supersymmetry. The Frontiers Of Quark - Lepton Physics,}
  (Springer, New York, 2003), p. 421.




%
%
%


%
%




\bibitem{ArkaniHamed:1998pf}
  N.~Arkani-Hamed and Y.~Grossman,
   Phys.\ Lett.\ B {\bf 459}, 179 (1999). 




\bibitem{Abazajian:2012ys}
  K.~N.~Abazajian {\it et al.},
   arXiv:1204.5379 [hep-ph];
  B.~Kayser,
  arXiv:1207.2167 [hep-ph].



\bibitem{interm-unif-gut}
 J.M.~Gipson and R.E.~Marshak,
  Phys.\ Rev.\ D {\bf 31}, 1705 (1985);
D.~Chang, R.N.~Mohapatra, J.M.~Gipson, R.E.~Marshak, M.K.~Parida,
  Phys.\ Rev.\ D {\bf 31}, 1718 (1985);
  N.G.~Deshpande, E.~Keith, P.B.~Pal,
   Phys.\ Rev.\ D {\bf 46}, 2261 (1992);
   J.L. Chkareuli, I.G. Gogoladze, A.B. Kobakhidze,
  Phys.\ Lett.\ B {\bf 340}, 63 (1994);
B.~Stech and Z.~Tavartkiladze,
   Phys.\ Rev.\ D {\bf 70}, 035002 (2004); 
   Phys.\ Rev.\ D {\bf 77}, 076009 (2008); 
  S.~Bertolini, L.~Di Luzio, M.~Malinsky,
   Phys.\ Rev.\ D {\bf 80}, 015013 (2009); 
K.S.~Babu and R.N.~Mohapatra,
   Phys.\ Lett.\ B {\bf 715}, 328 (2012). 



\bibitem{d6-flavored}
 A.~De Rujula, H.~Georgi and S.~L.~Glashow,
   Phys.\ Rev.\ Lett.\  {\bf 45}, 413 (1980);
M.~Yoshimura,
   Prog.\ Theor.\ Phys.\  {\bf 64}, 1756 (1980);
  D.~B.~Reiss and S.~Rudaz,
   Phys.\ Rev.\ D {\bf 30}, 118 (1984).


\bibitem{d6-rotate}
C.~Jarlskog,
 Phys.\ Lett.\ B {\bf 82}, 401 (1979);
R.N.~Mohapatra,
    Phys.\ Rev.\ Lett.\  {\bf 43}, 893 (1979);
  S.~Nandi, A.~Stern and E.C.G.~Sudarshan,
  Phys.\ Lett.\ B {\bf 113}, 165 (1982);
  Q.~Shafi and Z.~Tavartkiladze,
  Phys.\ Lett.\ B {\bf 451}, 129 (1999);
   Nucl.\ Phys.\ B {\bf 573}, 40 (2000);
  I.~Dorsner and P.~Fileviez Perez,
    Phys.\ Lett.\ B {\bf 605}, 391 (2005);
   Phys.\ Lett.\ B {\bf 606}, 367 (2005);
  Z.~Tavartkiladze,
    Phys.\ Rev.\ D {\bf 76}, 055012 (2007);
K.~-S.~Choi,
  Phys.\ Lett.\ B {\bf 668}, 392 (2008). 




\bibitem{Beringer:1900zz}
  J.~Beringer {\it et al.}  [Particle Data Group Collaboration],
  ``Review of Particle Physics (RPP),''  Phys.\ Rev.\ D {\bf 86}, 010001 (2012).

\bibitem{p-decay-widths}
M. Claudson, M.B. Wise and L.J. Hall,
    Nucl.\ Phys.\ B {\bf 195}, 297 (1982);
    S. Chadha and M. Daniel,
  Nucl.\ Phys.\ B {\bf 229}, 105 (1983);
  for a useful summary of formulas for nucleon decay widths, see
  V. Lucas and S. Raby,
  Phys.\ Rev.\ D {\bf 55}, 6986 (1997).

\bibitem{Nihei:1994tx}
  T.~Nihei and J.~Arafune,
   Prog.\ Theor.\ Phys.\  {\bf 93}, 665 (1995). 

\bibitem{Buras:1977yy}
  A.~J.~Buras, J.~R.~Ellis, M.~K.~Gaillard and D.~V.~Nanopoulos,
    Nucl.\ Phys.\ B {\bf 135}, 66 (1978);
  J.~T.~Goldman and D.~A.~Ross,
 Nucl.\ Phys.\ B {\bf 171}, 273 (1980).


\bibitem{Babu:2013jba}
  K.~S.~Babu, E.~Kearns, U.~Al-Binni, {\it et al.},
 arXiv:1311.5285 [hep-ph];
S.~Raby, T.~Walker, K.~S.~Babu, {\it et al.},
    arXiv:0810.4551 [hep-ph].

\bibitem{Ade:2014xna}
  P.~A.~R.~Ade {\it et al.}  [BICEP2 Collaboration],
   Phys.\ Rev.\ Lett.\  {\bf 112}, 241101 (2014). 

\bibitem{Hook:2014uia}
A.~Kobakhidze and A.~Spencer-Smith,
   Phys.\ Lett.\ B {\bf 722}, 130 (2013); 
   A.~Hook, J.~Kearney, B.~Shakya and K.~M.~Zurek,
   arXiv:1404.5953 [hep-ph].

   \bibitem{Espinosa:2007qp}
  J.~R.~Espinosa, G.~F.~Giudice and A.~Riotto,
   JCAP {\bf 0805}, 002 (2008);
   M.~Fairbairn and R.~Hogan,
   Phys.\ Rev.\ Lett.\  {\bf 112}, 201801 (2014); 
   K.~Enqvist, T.~Meriniemi and S.~Nurmi,
  arXiv:1404.3699 [hep-ph];
  A.~Kobakhidze and A.~Spencer-Smith,
   arXiv:1404.4709 [hep-ph].


\bibitem{Chakrabortty:2013mha}
  J.~Chakrabortty, P.~Konar and T.~Mondal,
   Phys.\ Rev.\ D {\bf 89}, 095008 (2014).


\bibitem{Passera:2010ev}
  M.~Passera, W.~J.~Marciano and A.~Sirlin,
  Chin.\ Phys.\ C {\bf 34}, 735 (2010)  [arXiv:1001.4528 [hep-ph]];
    AIP Conf.\ Proc.\  {\bf 1078}, 378 (2009)  [arXiv:0809.4062 [hep-ph]].

\bibitem{LeeRoberts:2011zz}
  B.~Lee Roberts [Fermilab P989 Collaboration],
   Nucl.\ Phys.\ Proc.\ Suppl.\  {\bf 218}, 237 (2011);
   N.~Saito [J-PARC g-2/EDM Collaboration],
    AIP Conf.\ Proc.\  {\bf 1467}, 45 (2012).

\bibitem{Adam:2013mnn}
  J.~Adam {\it et al.}  [MEG Collaboration],
   Phys.\ Rev.\ Lett.\  {\bf 110}, 201801 (2013). 



%
%

\bibitem{Carone:2012cd}
C.~D.~Carone and M.~Golden,
   Phys.\ Rev.\ D {\bf 49}, 6211 (1994); 
  C.~D.~Carone,
    Phys.\ Rev.\ D {\bf 86}, 055011 (2012); 
  S.~Bar-Shalom,
   arXiv:1310.2942 [hep-ph];
  T.~Abe and R.~Kitano,
  Phys.\ Rev.\ D {\bf 88}, 015019 (2013). 


\bibitem{Coleman:1980mx}
  S.~R.~Coleman and E.~Witten,
   Phys.\ Rev.\ Lett.\  {\bf 45}, 100 (1980).



\bibitem{Appelquist:2009ka}
  T.~Appelquist {\it et al.}  [LSD Collaboration],
   Phys.\ Rev.\ Lett.\  {\bf 104}, 071601 (2010); 
  T. Appelquist and Y. Bai,
   Phys.\ Rev.\ D {\bf 82}, 071701 (2010); 
  T. Appelquist, J. Terning and L.C.R. Wijewardhana,
   Phys.\ Rev.\ Lett.\  {\bf 79}, 2767 (1997); 
  T. Appelquist, Z.-y. Duan and F. Sannino,
 Phys.\ Rev.\ D {\bf 61}, 125009 (2000); 
  D. Areán, I. Iatrakis, M. Järvinen and E. Kiritsis,
   JHEP {\bf 1311}, 068 (2013). 





\bibitem{DelDebbio:2010zz}
  L.~Del Debbio,
   PoS LATTICE {\bf 2010}, 004 (2010)  [arXiv:1102.4066 [hep-lat]];
  M.~Piai,
   Adv.\ High Energy Phys.\  {\bf 2010}, 464302 (2010);  
   and references therein.




\bibitem{Belanger:2013xza}
  G.~Belanger, B.~Dumont, U.~Ellwanger, J.~F.~Gunion and S.~Kraml,
   Phys.\ Rev.\ D {\bf 88}, 075008 (2013); 
A.~Djouadi and G.~ég.~Moreau,
   arXiv:1303.6591 [hep-ph];
  P.~Bechtle, S.~Heinemeyer, O.~Stål, T.~Stefaniak and G.~Weiglein,
    arXiv:1305.1933 [hep-ph].




\bibitem{Eichten:1983hw}
  E.J.~Eichten, K.~D.~Lane and M.~E.~Peskin,
  Phys.\ Rev.\ Lett.\  {\bf 50}, 811 (1983).

\bibitem{Hod:2013cba}
  N.~T.~Hod [ATLAS Collaboration],
   EPJ Web Conf.\  {\bf 49}, 15004 (2013); 
  G.~Aad {\it et al.}  [ATLAS Collaboration],
  JHEP {\bf 1211}, 138 (2012); 
  S.~Chatrchyan {\it et al.}  [CMS Collaboration],
   Phys.\ Rev.\ Lett.\  {\bf 109}, 141801 (2012); 
    and references therein.




\bibitem{Eichten:2012hs}
  E.~Eichten, K.~Lane, A.~Martin and E.~Pilon,
   Phys.\ Rev.\ D {\bf 86}, 074015 (2012);
   arXiv:1201.4396 [hep-ph].


%
%

\bibitem{Einhorn:1981cy}
M.~J.~G.~Veltman,
   Nucl.\ Phys.\ B {\bf 123}, 89 (1977);
  M.~B.~Einhorn, D.~R.~T.~Jones and M.~J.~G.~Veltman,
   Nucl.\ Phys.\ B {\bf 191}, 146 (1981);
   M.~E.~Peskin and T.~Takeuchi,
  Phys.\ Rev.\ Lett.\  {\bf 65}, 964 (1990);
  H.-J.~He, N.~Polonsky and S.-f.~Su,
  Phys.\ Rev.\ D {\bf 64}, 053004 (2001). 


\bibitem{Peskin:1991sw}
  M.~E.~Peskin and T.~Takeuchi,
   Phys.\ Rev.\ D {\bf 46}, 381 (1992);
   A.~Orgogozo and S.~Rychkov,
   JHEP {\bf 1203}, 046 (2012); 
    JHEP {\bf 1306}, 014 (2013).

\bibitem{Lavoura:1992np}
  L.~Lavoura and J.~P.~Silva,
    Phys.\ Rev.\ D {\bf 47}, 2046 (1993);
    G.~Cynolter and E.~Lendvai,
   Eur.\ Phys.\ J.\ C {\bf 58}, 463 (2008); 
   S.~Dawson and E.~Furlan,
   Phys.\ Rev.\ D {\bf 86}, 015021 (2012). 


\bibitem{Baak:2011ze}
  M.~Baak, M.~Goebel, J.~Haller, A.~Hoecker, D.~Ludwig, K.~Moenig, M.~Schott and J.~Stelzer,
  Eur.\ Phys.\ J.\ C {\bf 72}, 2003 (2012). 

\bibitem{Dvali:1995cc}
P.~Salomonson, B.~S.~Skagerstam and A.~Stern,
  Phys.\ Lett.\ B {\bf 151}, 243 (1985);
  G.~R.~Dvali and G.~Senjanovic,
   Phys.\ Rev.\ Lett.\  {\bf 74}, 5178 (1995); 
G.~R.~Dvali, A.~Melfo and G.~Senjanovic,
  Phys.\ Rev.\ Lett.\  {\bf 75}, 4559 (1995). 


\bibitem{Langacker:1980kd}
  P.~Langacker and S.~-Y.~Pi,
   Phys.\ Rev.\ Lett.\  {\bf 45}, 1 (1980);
V.~V.~Dixit and M.~Sher,
   Phys.\ Rev.\ Lett.\  {\bf 68}, 560 (1992);
   T.~H.~Farris, T.~W.~Kephart, T.~J.~Weiler and T.~C.~Yuan,
    Phys.\ Rev.\ Lett.\  {\bf 68}, 564 (1992);
    G.~R.~Dvali, H.~Liu and T.~Vachaspati,
   Phys.\ Rev.\ Lett.\  {\bf 80}, 2281 (1998); 
   B.~Bajc, A.~Riotto and G.~Senjanovic,
   Phys.\ Rev.\ Lett.\  {\bf 81}, 1355 (1998) 
   and references therein.


\bibitem{Slansky:1981yr}
  R.~Slansky,
   Phys.\ Rept.\  {\bf 79}, 1 (1981).


%
%

\bibitem{jonesRG}
  D.~R.~T.~Jones,
  Phys.\ Rev.\ D {\bf 25}, 581 (1982);
  M.~E.~Machacek, M.~T.~Vaughn,
   Nucl.\ Phys.\ B {\bf 222}, 83 (1983);
  Nucl.\ Phys.\ B {\bf 236}, 221 (1984).



\bibitem{Langacker:1992rq}
  P.~Langacker and N.~Polonsky,
  Phys.\ Rev.\ D {\bf 47}, 4028 (1993).

%
%

\bibitem{Holdom:1985ag}
  B.~Holdom,
   Phys.\ Lett.\ B {\bf 166}, 196 (1986).

   \bibitem{Babu:1996vt}
  K.~S.~Babu, C.~F.~Kolda and J.~March-Russell,
   Phys.\ Rev.\ D {\bf 54}, 4635 (1996). 

\end{thebibliography}

\end{document}